\documentclass[nofootinbib, reprint, floatfix]{revtex4-2}
\usepackage{float}
\usepackage{upgreek}
\usepackage{siunitx}
\usepackage{amsmath,amsfonts, amssymb,graphicx,amsthm,color}
\usepackage{mathtools}
\usepackage{dcolumn}
\usepackage{bm}
\usepackage{rotating}
\usepackage{hyperref}
\usepackage{epstopdf}
\usepackage{algorithm}
\usepackage{array}
\usepackage{algpseudocode}
\newcolumntype{M}[1]{>{\raggedright}m{#1}}

\begin{document}

\title{Simulation of Rydberg Ionization in Atomic Beams for FIB Optimization}

\author{Cl\'elia Bastelica$^{1}$, Azer Trimeche$^{1}$, Colin Lopez$^{2}$, Matthieu Viteau$^{3}$, Patrick Cheinet$^{1}$, Daniel Comparat$^{1}$ and Yan J. Picard$^{1}$}
\email[Corresponding author: ]{yan.picard@universite-paris-saclay.fr}

\affiliation{
$^{1}$ Universit\'e Paris-Saclay, CNRS, Laboratoire Aim\'e Cotton, 91405 Orsay, France\\
$^{2}$ Universit\'e Paris-Saclay, CNRS, Ecole Normale Sup\'erieure Paris-Saclay, CentraleSup\'elec, LuMIn, 91405 Orsay, France\\
$^{3}$ Tescan, ZAC Saint Charles n$^{\circ}$95, 3$^e$ Avenue 13710 Fuveau, France
}

\date{\today}

\begin{abstract}
\small
This study explores the excitation and ionization of an atomic beam as a pathway to optimize focused ion beams (FIBs) for high-precision applications. Leveraging the unique advantages of Rydberg excitation followed by field ionization---specifically its ability to minimize velocity and position dispersions---we present a method to generate ion beams with good performance at low energies. A custom Lua program, integrated into the SIMION simulation platform, models the intricate processes of particle distributions, laser excitation, and Rydberg ionization. This integrated approach incorporates essential parameters such as excitation and ionization rates, Stark shifts, Doppler effects, and electric fields, enabling a detailed analysis of ion beam properties. Our simulations demonstrate the influence of critical factors such as the chosen Rydberg state, ionization region characteristics, and velocity dispersions on the final ion beam quality. By optimizing these parameters, we achieve significant reduction of the axial energy spread and the longitudinal extent of the ionization region. This framework bridges theoretical modeling and experimental validation, offering a comprehensive toolkit for the development of next-generation ion sources and advancing FIB technologies across various scientific domains.
 \end{abstract}

\maketitle

\section{Introduction}

Ion and electron sources are indispensable tools in numerous fields, ranging from spectroscopy and quantum technologies to nanotechnology and materials science \cite{Materials2019,laucht2021roadmap,hoflich2023roadmap}. The high-precision generation, manipulation and control of beams of charged particles underpin various cutting-edge applications, including high-resolution imaging, ion beam lithography, nanofabrication, particle acceleration, and ion-beam-based cancer therapy. As experimental demands grow, the development of innovative methods for particle generation and control has become increasingly critical \cite{hawkes2019springer,li2021recent}.

Complementary to the various approaches to generate charged particles, such as thermionic, field emission, or photoemission guns for electrons and plasma-based, liquid metal ion sources (LMIS), or gas field ion sources (GFIS) for FIBs, the use of (potentially laser-cooled) neutral atomic beam ionization stands out as a powerful technique for producing bright and monochromatic electron and ion beams \cite{Gallagher.1974,musumeci2018advances,hoflich2023roadmap}. It is particularly useful
for applications that require high spatial resolution, such as FIB nanofabrication \cite{hoflich2023roadmap}. Several studies, including our own, have investigated various ionization strategies in this context \cite{mcclelland2016bright,mcculloch2016cold}. Among them, excitation to Rydberg states followed by electric-field ionization --- hereafter referred to as Rydberg ionization --- offers distinctive control over the ionization pathway by promoting atoms to high-principal-quantum-number states before field-induced ionization \cite{gallagher1994rydberg}. This approach enables the generation of ions or electrons with exceptional energy precision and reproducibility.

\begin{figure}
    \centering
    \includegraphics[width=\linewidth]{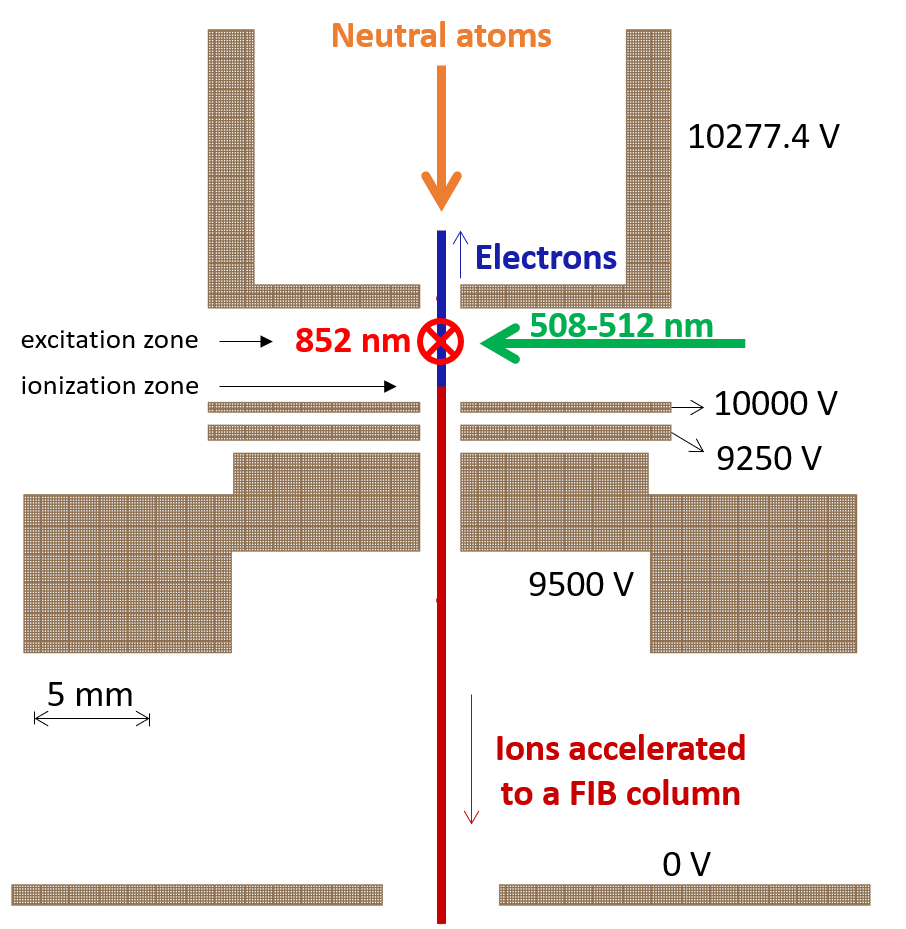}
    \caption{
    \textsc{SIMION} schematic of the ion source in the ColdFIB setup, showing key parameters such as electrode voltages, electrode geometry, and the wavelengths and orientations of the excitation lasers. The waists of both excitation beams are identical, with $W_0 = 50~\mu\mathrm{m}$. The remaining elements of the FIB column are described in Figure~\ref{Coldfib}; here we focus on the excitation and ionization regions. Neutral cesium atoms (orange arrow) emerge from an oven in the ground state and are excited to a Rydberg state via a two-photon scheme driven by the red and green lasers. Upon entering the region of resonant electric field, these Rydberg atoms are field-ionized, producing ions (dark red) and electrons (blue) that are extracted in opposite directions. In the present configuration, only the ions are collected and subsequently accelerated and focused by the FIB column.}
    \label{exemple_FIB_simion}
\end{figure}

In our group, we develop customized advanced ion and electron sources to support a wide range of studies and applications, including precision spectroscopy, nanofabrication, and quantum experiments \cite{hahn2021cesium,trimeche2020ion,lopez2019real,mcculloch2017field,gallagher1994rydberg,viteau2016ion,antoni2018watt,reveillard2018coldfib,hahn2022comparative}. To optimize source designs and interpret experimental results, we use \textsc{SIMION} \cite{manura20088} to compute electrostatic fields and charged-particle trajectories in realistic electrode geometries. The neutral-to-charged conversion, however, is not an empirical source model here: it is set by experimentally controlled parameters (laser geometry and intensity profiles, atomic-beam velocity distribution, and the field-dependent excitation/ionization dynamics). In the following, we present a computational framework that combines these ingredients and apply it to the ColdFIB configuration \cite{viteau2016ion} (Figures~\ref{exemple_FIB_simion} and \ref{Coldfib}).

Figure~\ref{exemple_FIB_simion} shows the ion-source geometry as implemented in \textsc{SIMION}, together with representative trajectories of the incoming neutral cesium atoms. In conventional FIB simulations, the definition of the ion (or electron) creation volume and the corresponding initial phase-space distribution is a critical and often empirical modeling choice. In our case, these quantities are instead set by experimentally controlled parameters: the positions and waists of the excitation lasers and the temperature of the oven producing the neutral cesium beam. These parameters fully determine the Rydberg excitation process and thus provide precise control over ionization. Cesium atoms are resonantly driven from the ground state $6S_{1/2},F=4$ to the intermediate state $6P_{3/2},F=5$, and then to the selected Rydberg state. The lasers used for these two excitation steps are indicated in red and green, respectively, in Figure ~\ref{exemple_FIB_simion}.

Given the complexity of excitation and ionization processes, we extend \textsc{SIMION} with a Lua module that implements the internal-state stochastic dynamics (laser-driven excitation, spontaneous decay, and field-ionization rates) along each particle trajectory. The model evaluates excitation and ionization rates, including Doppler and Stark shifts, yielding physically constrained ion-creation volumes and initial-energy distributions that can be propagated through the column for quantitative comparison with measurements and for source optimization. Using this framework, we analyze how the choice of Rydberg state, the relative placement of excitation and ionization zones, and the local electric-field gradient control the longitudinal ionization distribution and hence the extracted kinetic-energy dispersion that sets the dominant source contributions to geometric and chromatic aberrations in low-energy FIB operation. The simulations are confronted with experimental measurements from the ColdFIB setup, providing a validated basis for interpreting the data and optimizing the source, rather than presenting software for its own sake.

The article is organized as follows. We first present an overview of the computational framework developed to model the excitation and ionization processes, including the underlying physical principles and the calculation of excitation and ionization rates, Stark shifts, and Doppler effects. We then describe the implementation of these ingredients in our custom Lua program and its integration with \textsc{SIMION}, which enables a simultaneous treatment of the external degrees of freedom (particle dynamics in electric fields) and the internal quantum-state dynamics of the Rydberg atoms. Finally, we apply this methodology to the optimization of FIB sources, analyzing the interplay between ion-beam velocity dispersion, the spatial extent of the ionization region, and their combined impact on the extracted ion energy spread. Using experimental data from the ColdFIB setup to validate the simulations, we demonstrate the strong influence of excitation and ionization parameters on the achievable beam resolution. These results highlight the versatility and predictive power of our hybrid Lua-\textsc{SIMION} framework and point to applications in focused ion beams, electron and ion microscopy, and advanced quantum technologies.

\section{Simulation Framework for Excitation and Ionization}

The Lua program developed by our group fully simulates an atomic beam, initially excited by lasers to a Rydberg state and subsequently ionized by an electric field, as shown in Figure \ref{exemple_FIB_simion}. Rydberg ionization, which ionizes particles already excited to Rydberg states, is a highly efficient method for achieving controlled ionization with precise energy. This technique produces charged particles with significantly reduced velocity dispersion compared to conventional methods, such as the most used ion sources in FIBs, the Liquid Metal Ion Sources and the plasma sources.


One of the main challenges in this process is the direct excitation of neutral atoms from their ground state to the Rydberg states. Such transitions typically require lasers with specific wavelengths and extremely narrow linewidths, which are often difficult to achieve experimentally. To overcome this, multiphoton excitation schemes are commonly employed, in which several lower-energy photons are absorbed sequentially to reach the desired Rydberg state. This approach relaxes the laser requirements while maintaining the precision necessary for controlled Rydberg ionization. Furthermore, using multiple lasers allows spatial selection of the excitation region, and consequently, of the ionization region. 
In our case the cesium excitation process proceeds as follows:
\begin{enumerate}
    \item A 852 nm laser, locked at resonance, excites atoms from the $6S_{1/2, F=4}$ ground state $|0\rangle$ to the $6P_{3/2, F=5}$ intermediate state $|1\rangle$. This laser is oriented transversely to the atomic beam.
    \item A second laser, operating at 508--512 nm, further excites the atoms from $6P_{3/2, F=5}$ to a Rydberg state $|2\rangle$. This laser is also transverse to the beam and perpendicular to the first laser.  
\end{enumerate}
Experimentally, the efficiency of excitation to a Rydberg state depends on the laser polarization. In this work, this effect is taken into account implicitly: the laser polarization is assumed to be appropriate for driving the chosen transition.

The simulation framework we developed combines a custom Lua program with the \textsc{SIMION} environment to model excitation and ionization processes with high flexibility and accuracy. The workflow begins in \textsc{SIMION}, where key parameters of the atomic beam are defined, including the particle species and initial longitudinal positions. Particle velocities are then assigned either directly within \textsc{SIMION} or via the Lua program, depending on the chosen beam distribution, which is informed by experimental measurements.

All laser-related parameters used for excitation are specified in an external input file. This includes beam profiles, waists, powers, wavelengths, linewidths, propagation directions, and positions. Additionally, critical experimental parameters---such as electrode voltages, the spatial extent of excitation and ionization regions, and the number of simulated particles---are also defined in this file. User-defined values for atomic properties, including the electric dipole moment, transition frequencies, and excitation slopes, can be derived from theoretical calculations, experimental data, or set manually to suit the simulation requirements.

The Lua program reads these input files and generates the corresponding simulation environment in \textsc{SIMION}, automating the assignment of parameters, calculating time-dependent rates, and controlling the stochastic evolution of each particle. This modular approach ensures a fully configurable and adaptable framework that can easily accommodate a wide range of experimental setups. In the following sections, we detail the specific parameter choices implemented for the present study.

\begin{figure}
    \centering
    \includegraphics[width=\linewidth]{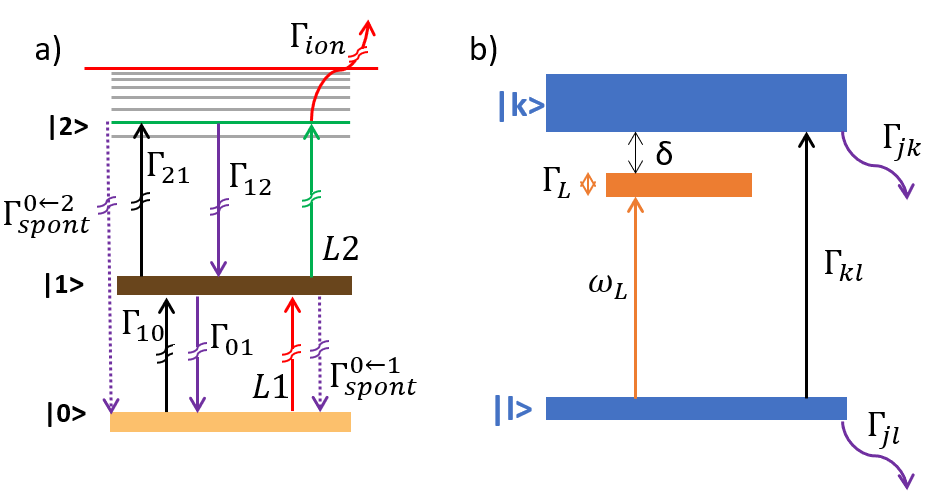}
    \caption{
    (a) Schematic of a two-photon Rydberg ionization process. The atom is excited from the ground state \(|0\rangle\) to the intermediate state \(|1\rangle\) at a rate \(\Gamma_{10}\) induced by laser \(L_1\). From \(|1\rangle\), the atom can either return to \(|0\rangle\) via stimulated emission at a rate \(\Gamma_{01} = \Gamma_{10}\) or decay spontaneously at a rate \( \Gamma_{\rm spont}^{0 \leftarrow 1} = 1/\tau\), where \(\tau\) is the lifetime of \(|1\rangle\). The same scheme applies to the subsequent transition to the Rydberg state. Atoms in the Rydberg state \(|2\rangle\) have a finite probability of ionization in the presence of an electric field, characterized by the rate \(\Gamma_{\rm ion}\). (b) Schematic of an open two-level system driven by a detuned laser. The rates \(\Gamma_{jk}\) and \(\Gamma_{jl}\) correspond to transitions from states \(|k\rangle\) and \(|l\rangle\), respectively, to another state \(|j\rangle\). Here, \(\Gamma_L\) denotes the laser linewidth (FWHM), \(\omega_L\) is the laser angular frequency, and \(\delta\) is the detuning between the laser and atomic transition angular frequencies.}
    \label{taux}
\end{figure}

Figure \ref{taux}(a) illustrates a multiphoton ionization process in which two Gaussian laser beams excite neutral atoms to a well-defined Rydberg state, followed by ionization induced by an external electric field. Figure \ref{taux}(b) provides a schematic of a two-level system, $|l\rangle$ and $|k\rangle$, highlighting the key parameters used to compute excitation and ionization rates. These parameters include transition rates, laser detunings, and linewidths, providing a comprehensive depiction of the physical processes governing atomic excitation and subsequent ionization.

To describe the time evolution of the population in a given state $|k\rangle$, we begin with the standard rate equation, following the notation in Figure \ref{taux}(b):

\begin{equation}
\frac{dN_k}{dt} = \sum_l N_l \Gamma_{k\leftarrow l} - N_k \sum_l \Gamma_{l \leftarrow k}  
\label{eq:population_dynamics_split} 
\end{equation}

where $N_l$ and $N_k$ denote the populations of states $|l\rangle$ and $|k\rangle$, respectively. Depending on the state $|k\rangle$ under consideration, the rates $ \Gamma_{k\leftarrow l}$, which express the gain to state $k$, and $\Gamma_{l \leftarrow k}$, which express the loss from it, are combinations of the following: 
\begin{itemize}
    \item \( \Gamma_{lk}  \): stimulated transition rate from state \(|k\rangle\) to \(|l\rangle\), including excitation or stimulated emission that have the same rates so $\Gamma_{lk} = \Gamma_{kl}$.
    \item \( \Gamma_{\rm spont}^{l \leftarrow k} \): spontaneous decay rate from state \(|k\rangle\) to state $|l\rangle$. In our simplified model, they only go to the ground state \(|l=0\rangle\), all other rates are assumed to be zero.
    \item \( \Gamma_{\rm ion}^{k} \): ionization rate from state \(|k\rangle\), which is zero for the ground and intermediate states and equals \(\Gamma_{\rm ion}= \Gamma_{\rm ion,F} + \Gamma_{\rm ion,L}\) for the Rydberg states. Here, \(\Gamma_{\rm ion,F}\) denotes electric-field ionization in the extraction/ionization region, whereas \(\Gamma_{\rm ion,L}\) denotes direct photoionization by the excitation lasers.
\end{itemize}

Thus, the equation describes the dynamics of excitation, stimulated emission, spontaneous decay, and ionization processes, for an atom in a given state $|k\rangle$,  with $  \Gamma_{l\leftarrow k} = \Gamma_{lk}  + \Gamma_{\rm spont}^{0 \leftarrow k} + \Gamma_{\rm ion}^{k}$. 
Note that \(|l\rangle\) can correspond to either a lower or higher energy level, independent of its depiction as a lower state in Figure~\ref{taux}(b). In principle, Rydberg atoms may also undergo direct photoionization in the laser-interaction region  (see, e.g., \cite{viray2021photoionization}). However, for the laser parameters and atomic velocities considered in this work, this contribution is negligible compared to the intended electric-field ionization. We therefore set \(\Gamma_{\rm ion}\simeq \Gamma_{\rm ion,F}\) in the model and simulations presented below.

By applying certain approximations, such as the low-saturation regime and the rotating wave approximation (see Ref. \cite{comparat2014molecular}), the excitation rate $\Gamma_{lk}$ for a transition from state $|k\rangle$ to $|l\rangle$ can be expressed as:
\begin{equation}
\Gamma_{lk} = \Omega^2_{lk} \frac{\Gamma_{\rm tot}}{\Gamma_{\rm tot}^2 + 4 \delta^2}
\label{gamma_ij}
\end{equation}
where $\Omega_{lk}$ is the Rabi frequency of the transition, $\Gamma_{\rm tot}$ is the total decay rate, including all incoherent transition rates from \(|l\rangle\) and \(|k\rangle\), and $\delta$ is the detuning between the laser frequency \(\omega_L\) and the atomic transition frequency \(\omega_{lk}\). In Eq.~\eqref{gamma_ij}, the total decay rate $\Gamma_{\rm tot}$ is expressed as:
\begin{align*}
\Gamma_{\rm tot} = \Gamma_L  + \sum_{j\neq k}  \Gamma_{jl}  + \sum_{j} \Gamma_{\rm spont}^{j\leftarrow l} + \Gamma_{\rm ion}^{l} \\
+\sum_{j\neq l}  \Gamma_{jk} + \sum_j\Gamma_{\rm spont}^{j\leftarrow k}+ \Gamma_{\rm ion}^{k}
\end{align*}

where $\Gamma_L$ is the full-width of the laser at half-maximum (FWHM) linewidth. In our calculations, the Rabi frequency is given by:
\begin{equation}
\Omega_{lk}^2 = \frac{2 d_{lk}^2 I_L}{\epsilon_0 c \hbar^2}
\end{equation}
where $d_{lk}$ is the electric dipole moment of the transition, and $I_L$ denotes the laser intensity, evaluated at each position for a Gaussian laser beam using the specified laser parameters. The laser frequency detuning $\delta$, denoted $\delta^L_{lk}$ for intermediate transitions, is defined as:
\begin{equation}
\delta^L_{lk} = \omega_L - \omega_{lk} - \mathbf{k} \cdot \mathbf{v}
\label{delta_lk}
\end{equation}
where $\mathbf{k}$ is the laser wavevector and the term $\mathbf{k} \cdot \mathbf{v}$ accounts for the Doppler shift due to the atomic motion along the laser propagation direction.

For transitions to the Rydberg state, the presence of an external electric field, used for the extraction of charged particles, introduces an additional dependence of detuning $\delta$ due to the Stark effect. Although the induced energy-level shift is negligible for intermediate transitions, it becomes significant for the Rydberg state. In this case, the detuning is expressed as a function of the local electric field $F$. In our Lua program, a first-order approximation assumes that the difference between the laser frequency $\omega_L$ and the atomic transition frequency $\omega_{lk}$ varies linearly with the electric field. For the Rydberg transition (from state $|1\rangle$ to state $|2\rangle$ in Figure \ref{taux}), the detuning, denoted $\delta_{\rm Ryd}$, is thus given by:
\begin{equation}
\delta_{\rm Ryd} = \alpha \, (F_0 - F) - \mathbf{k} \cdot \mathbf{v}
\label{deltaRyd}
\end{equation}
where $F$ is the electric field computed by \textsc{SIMION} at each position and $F_0$ is the resonant electric field, defined as the value at which the energy difference between the two levels matches the laser frequency. The parameter $\alpha$, representing the excitation slope, is extracted from the Stark diagram for the specific Rydberg state \cite{moufarej2017forced,chardonnet1984generation}.
Both $\alpha$ and $F_0$ are user-defined in the input file prior to simulation.

In this study, the excitation rates $\Gamma_{lk}$ for all transitions are computed at each spatial and temporal step using Eq.~\eqref{gamma_ij}. 
The ionization process follows a two-step mechanism: atoms are first excited to a Rydberg state via laser interactions and subsequently ionized by the applied electric field. This approach is particularly relevant for focused ion beam (FIB) applications, which require high ion flux --- critical for milling and etching --- while minimizing both positional and energy dispersion. Reducing these dispersions is essential to mitigate chromatic aberrations and achieve the smallest possible focal spot. To ensure a well-defined ionization region, the choice of Rydberg state, electrode geometry, and applied voltages is optimized so that the ionization field $F_{\rm ion}$ coincides with a region of strong electric field gradient. This configuration reduces the spread in both energy and position, thereby minimizing geometric and chromatic aberrations during focusing and enabling the smallest achievable beam size at the nanometer scale.

For singly charged ions, the kinetic-energy variation, $\Delta E$,  associated with an axial ionization-position spread \(\Delta x_{\rm ion}\) in an extraction field \(F\) is approximately \(\Delta E \approx F\,\Delta x_{\rm ion}\) (in eV unit). This estimate assumes a locally monotonic extraction potential, so that the final kinetic energy maps approximately linearly to the ionization position over the creation region. We define \(\Delta x_{\rm ion,FWHM}\) as the full width at half maximum of the axial distribution of ionization (creation) events along the propagation axis \(x\), evaluated at the instant of ionization. Since ions are produced inside the lenses responsible for extraction and initial focus before the condenser, longitudinal dispersion $\Delta x_{\rm ion}$ also contributes to geometric aberrations. These aberrations, distinct from those related to energy dispersion, are particularly enhanced in regions where the electric field lines deviate from being planar along the direction of propagation.

\begin{figure}
    \centering
    \includegraphics[width=\linewidth]{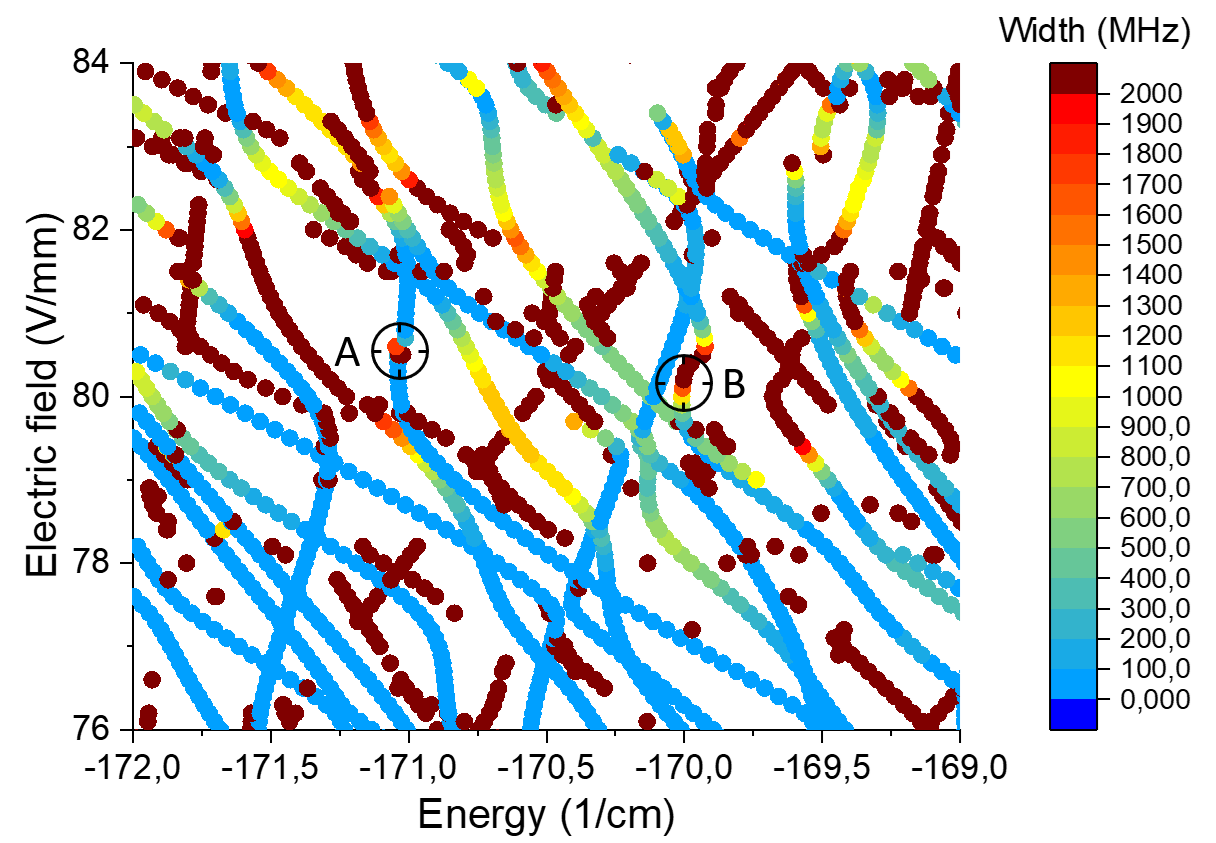}
    \caption{
    Photoabsorption Stark maps of cesium Rydberg states over a zero field binding energy range from $-172\,\mathrm{cm}^{-1}$ to $-169\,\mathrm{cm}^{-1}$ and an electric field varying from $75\,\mathrm{V/mm}$ to $85\,\mathrm{V/mm}$. The maps are computed using the local-frame transformation (LFT) theory at discrete electric field values, providing a computationally efficient framework for calculating the photoabsorption spectra of alkali-metal atoms \cite{hahn2021cesium}.
    The data points are color-coded to represent the linewidth (in MHz) of the Rydberg-state energy levels as a function of the applied electric field. This visualization makes it possible to track how individual Rydberg states evolve when approaching the ionization region (around $80~\mathrm V/mm$), providing simultaneous information on both their energy positions and ionization rates. Zones A and B highlight two regions with markedly different ionization behaviors: in zone A, the rapid variation of the Rydberg-state lifetime reflects a steeper local Rydberg Stark slope, whereas in zone B the slope is more shallow, resulting in a slower change of the ionization rate.
    }
    \label{ionization_rate}
\end{figure}

When the excitation and ionization regions are spatially separated, atoms promoted to a given Rydberg state traverse regions of varying electric field. Along this path, the chosen Rydberg level can undergo avoided crossings with nearby states before reaching the ionization zone, which may result in state mixing (see Figure~\ref{ionization_rate}). A quantitatively accurate treatment of these transitions would require a detailed multilevel calculation of the Stark dynamics, which lies beyond the scope of the present work.

When multiple crossings of energy levels occur, the wavefunctions of the involved states can mix, resulting in interference effects and multiple ionization thresholds. This wavefunction mixing can offer a means to control ionization, especially in cases of isolated two-level crossings where a stable Rydberg state couples to an unstable state. In such scenarios, the ionization rate (or lifetime) undergoes a sharp change. Ionization rates can be estimated through theoretical models \cite{hahn2021cesium,moufarej2017forced,mcculloch2017field}, spectroscopic measurements, or Monte Carlo simulations that account for excitation and ionization dynamics. These simulations typically utilize Stark maps computed using the local-frame transformation (LFT) theory for cesium Rydberg states near the classical field-ionization threshold \cite{hahn2021cesium}. The LFT theory provides the field-dependent photoabsorption cross-section \cite{Harmin1984PRA, hahn2021cesium}, from which the data presented in Figure \ref{ionization_rate} are extracted. Using the Stark diagram, we can precisely determine the experimental ionization rates for different Rydberg states, as demonstrated in previous studies \cite{hahn2021cesium, moufarej2017forced}. These theoretically derived rates are incorporated into the simulation to model the ionization behavior accurately. While the theory is complex, we simplify the modeling by using an exponential function \cite{moufarej2017forced} to approximate the ionization rate curves $\Gamma_{\rm ion}$ under these conditions:
\begin{equation}
\Gamma_{\rm ion} = \Gamma_{0} \, \exp\left(\frac{F - F_{\rm ion}}{\sigma}\right).
\label{gamma_ion_exp}
\end{equation}
In this framework, $\Gamma_0$ represents the ionization rate of a given Rydberg state near the threshold electric field $F_{\rm ion}$, while $\sigma$ characterizes the local Rydberg Stark slope in this region. These parameters define the ionization zone for Rydberg atoms. Within the Lua input file, $\Gamma_0$, $F_{\rm ion}$, and $\sigma$ are user-defined. At the same time, the local electric field $F$ is computed by SIMION at each spatial and temporal step during the simulation. These values depend on the selected Rydberg state and its evolution under the applied electric field. 

To validate the efficiency of the mathematical framework implemented in our code, we adapted the parameters in Eq.~\eqref{gamma_ion_exp} to compute the ionization rate for each Rydberg state. As shown in Figure \ref{ionization_rate} (examples A and B), the ionization efficiency---specifically, the rate at which Rydberg ionization occurs---strongly depends on the chosen atomic Rydberg state. Selecting an appropriate Rydberg state is crucial for each application, as it dictates the location and extent of the ionization region for given parameters (e.g., $F_{\rm ion}$, $\omega_L$, etc.). This choice directly influences both the spatial precision and the temporal dynamics of the ionization process.

After computing the excitation, de-excitation, and ionization rates for a given set of parameters, the next state of a particle---whether it transitions to a new state or remains in its current state $|k\rangle$---is determined. The core of the method involves solving the time evolution of the probability $P_k$ for the system to occupy state $k$, based on the master equation (cf. Eq.~\eqref{eq:population_dynamics_split}), written in probabilistic form as:
\begin{equation}
\frac{dP_k}{dt} = \sum_{l} \Gamma_{k\leftarrow l} P_l - \sum_{l} \Gamma_{l\leftarrow k} P_k .
\label{mastereq_main}
\end{equation}

Details of both existing algorithms and the one we derived are provided in Appendix \ref{KMC_algo}. Here, we briefly outline the main aspects. In our work we have used two different algorithms.
We first have used the simple stochastic  fixed time-step Monte Carlo method based on Euler's explicit discretization \cite{ceccatto1986effective}:
\[
P_k(t+\Delta t) \approx P_k(t) + \sum_l [\Gamma_{k\leftarrow l} P_l - \Gamma_{l\leftarrow k} P_k]\Delta t .
\]
However, to maintain reasonable accuracy, this method requires $\Gamma_k \Delta t \ll 1$, where $\Gamma_k = \sum_l \Gamma_{l\leftarrow k}$, leading to very small time steps $\Delta t$, which can significantly slow down the simulation. To overcome this limitation, we implement an alternative algorithm based on the kinetic Monte Carlo (KMC) method, which in our conditions is more than an order of magnitude faster than the standard Euler approach while maintaining equivalent accuracy. In this approach, the next reaction time $t'$ --- at which the system transitions to a new state $|l\rangle$ --- is determined directly. This is achieved by utilizing the fact that the expression 
$
e^{-\int_t^{t'} \Gamma_k(\tau)\,d\tau}
$
follows a uniform random distribution in the interval $]0,1]$, enabling efficient calculation of the reaction time $t'$ and reducing the computational cost.

This approach is statistically exact and, in principle, requires only a single time step. However, this holds strictly only for constant $\Gamma_k$, for which $t' = t - \ln(u)/\Gamma_k$, where $u$ is drawn uniform random number in the $(0,1]$ interval. When the total transition rate is time-dependent, $\Gamma_k=\Gamma_k(t)$, the integral must be approximated \cite{prados1997dynamical}. As detailed in the appendix, we derived a modified algorithm that constrains the time step $\Delta t$ to be smaller than $\epsilon \,\dfrac{\Gamma_k}{|d\Gamma_{k}/dt|}$, ensuring that the integral approximation has a controlled relative error $\epsilon$, i.e.,
$
\left| \int_t^{t+\Delta t} \,  \Gamma_{ k}(\tau)\, d\tau  - \Delta t \Gamma_{ k}(t) \right|  \leq \epsilon \Delta t \Gamma_{ k}(t)
$
which is valid provided the rates evolve smoothly. As demonstrated in the appendix, for Gaussian, Lorentzian, or exponential evolutions, $\epsilon < 0.3$ suffices. We thus choose a more conservative choice of $\epsilon < 0.1$ that ensures accurate evolution for nearly all physically reasonable rate variations. This guarantees controlled error for the evolution of the internal states. Simultaneously, the system evolves dynamically under the external timing $dt_{\rm SIMION}$ determined by the trajectory calculations. To integrate the internal (state) and external (trajectory) evolutions, the algorithm uses the smaller of the time steps at each iteration.

Our final Stochastic Reaction Time Evolution algorithm, termed Time-Dependent Kinetic Monte Carlo (TDKMC), is summarized as follows:
\begin{center}
\textbf{Time Dependent Kinetic Monte Carlo \\
(control Parameter $0<\epsilon<0.1$)}
\begin{algorithmic}[1]

\State $t \gets t_0$, $i \gets 0$, $I \gets 0$
\State Draw $u \sim \mathcal{U}(0,1]$
\State Initial state is named $|k\rangle$

\While{System stays in state $|k\rangle$}
  \State Compute rates $\Gamma_{l\leftarrow k}(t_i)$ for all states $|l\rangle$
  \State $\Gamma_k(t_i) \gets \sum_{l=1}^N \Gamma_{l\leftarrow k}(t_i)$
  \State $\Gamma'_{k,\text{approx}}(t_i) \gets 
    \begin{cases}
      \text{large value (safe)}, & i=0 \\
      \dfrac{\Gamma_k(t_i)-\Gamma_k(t_{i-1})}{t_i - t_{i-1}}, & i>0
    \end{cases}$

 \State \textbf{Compute candidate time step} $\Delta t_1$ (Reaction condition),  $\Delta t_2$ (error control),  $\Delta t_3$ (external step):
\If{$\Gamma_k(t_i) = 0$}
    \State $\Delta t \gets dt_{\text{SIMION}}$ \Comment{only external timing}
\Else
    \State $\Delta t_1 \gets \dfrac{-\ln u - I}{\Gamma_k(t_i)}$ 
    \State $\Delta t_2 \gets 
        \begin{cases}
            \text{large}, & \Gamma'_{k,\text{approx}}(t_i) = 0 \\
            \epsilon \,\dfrac{\Gamma_k(t_i)}{|\Gamma'_{k,\text{approx}}(t_i)|}, & \text{otherwise}
        \end{cases}$
    \State $\Delta t_3 \gets dt_{\text{SIMION}}$
    \State $\Delta t \gets \min(\Delta t_1, \Delta t_2, \Delta t_3)$
\EndIf

 \State Evolves external degree of freedom during $\Delta t $  (so forcing $dt_{\text{SIMION}}= \Delta t $)
  \State $t_{i+1} \gets t_i + \Delta t$

  \If{$\Delta t = \Delta t_1$}
    \State \textbf{Reaction occurs}
    \Statex \hspace{4em} Compute new rate list $\Gamma_{lk}(t_{i+1})$
    \Statex \hspace{4em} $R_j \gets \sum_{m=1}^j \Gamma_{mk}(t_{i+1})$, $R_0 \gets 0$
    \Statex \hspace{4em} Draw $u' \sim \mathcal{U}(0,1]$
    \Statex \hspace{4em} Find $l$ such that $R_{l-1} < u'R_N \leq R_l$
    \Statex \hspace{4em} Set system state $\gets |l\rangle$
    \State Restart from beginning (Line 1)
  \Else
    \State \textbf{No reaction}
    \Statex \hspace{4em} $I \gets I + \Delta t \cdot \Gamma_k(t_i)$
    \Statex \hspace{4em} $i \gets i+1$
    \State Continue loop (Line 5)
  \EndIf
\EndWhile

\end{algorithmic}

\end{center}

\begin{figure}[!h]
    \centering
    (a)
    \includegraphics[width=\linewidth]{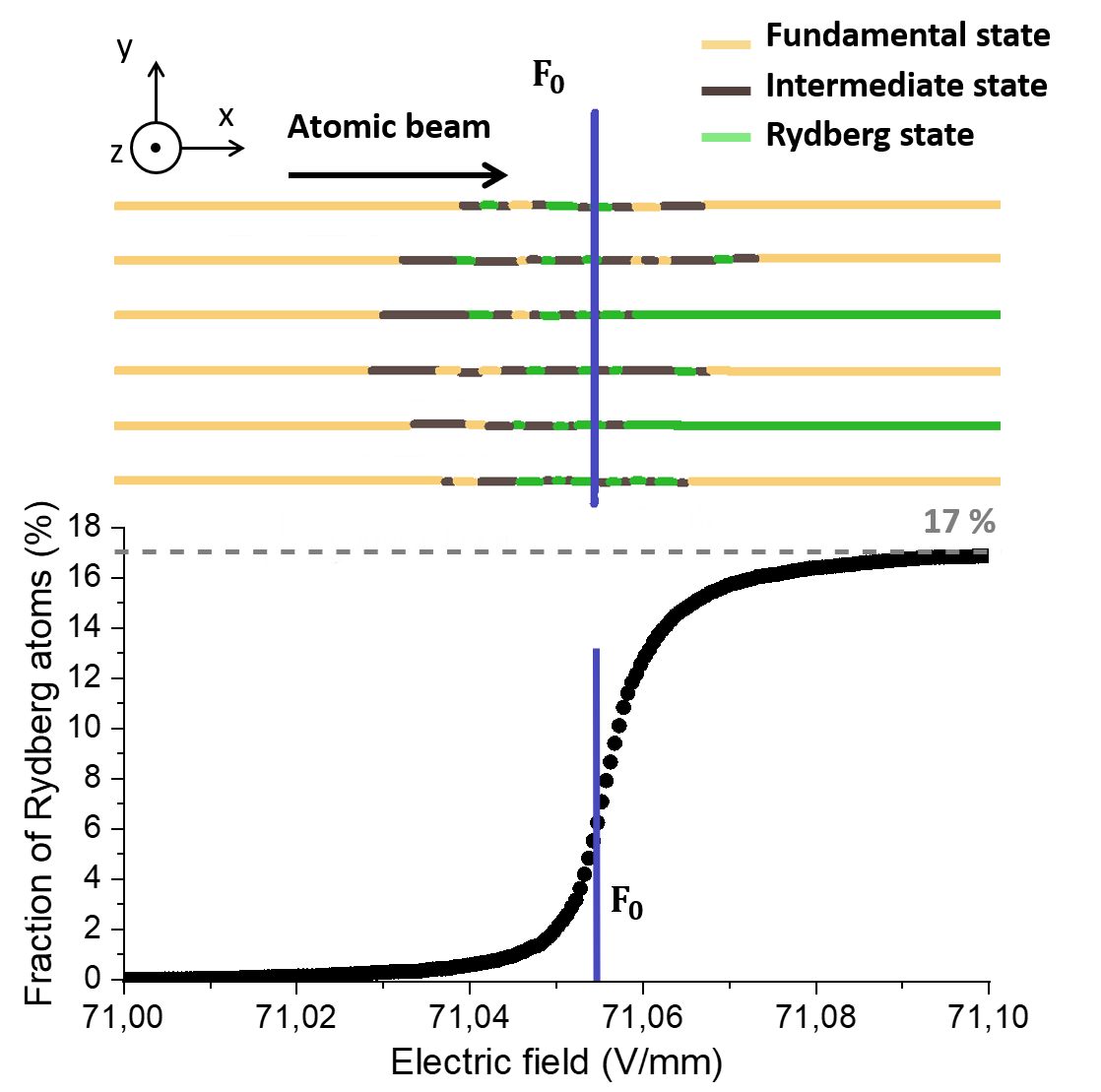}
    (b)
    \includegraphics[width=\linewidth]{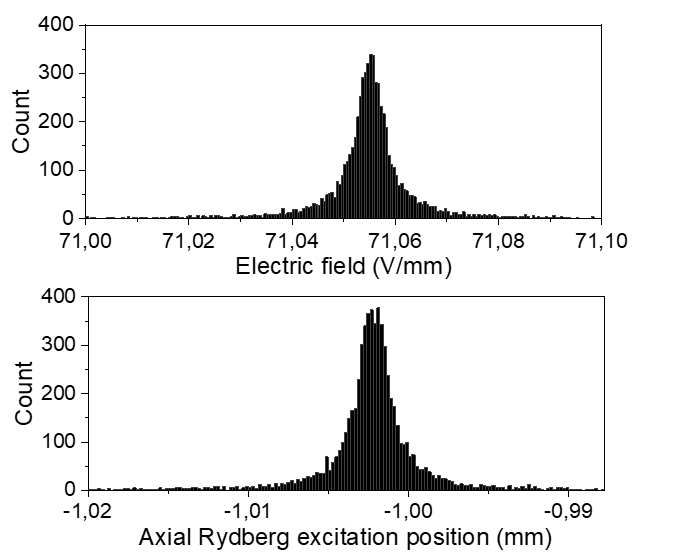}
    
    \caption{
            Simulation of the two-photon Rydberg excitation process in \textsc{SIMION} using our Lua-based program. (a) Top: representative trajectories of cesium atoms in the ground state (orange), intermediate state (dark brown), and Rydberg state (green). The parameter $F_0 = 71.056~\mathrm{V/mm}$ denotes the resonant excitation electric field. Color changes along the trajectories illustrate the stochastic excitation and de-excitation dynamics. Bottom: corresponding fraction of atoms in the Rydberg state as a function of the electric field. (b) Distribution of excitation events to the Rydberg state as a function of electric field (top) and axial position (bottom). All positions at which atoms are promoted to the Rydberg state are included, even when multiple de-excitation and re-excitation events occur along a single trajectory. This leads to Gaussian-like distributions centered around the resonant field $F_0$.
            }
    \label{excitation}
\end{figure}

This TDKMC approach provides an efficient framework for simulating the stochastic processes of excitation, de-excitation, and ionization of Rydberg atoms, while fully accounting for spatially varying electric fields and time-dependent transition rates. At each transition occurrence, the next state of a particle is determined based on the relevant transition probabilities. To ensure accurate tracking of the particle's evolution, each state is assigned a unique label during the simulation, allowing precise monitoring of its progression over time. This approach enables the simultaneous study of the trajectories of produced ions, electrons, or both, depending on the parameters defined in the simulation.

\begin{figure}[h!!]
    \centering
    \includegraphics[width=\linewidth]{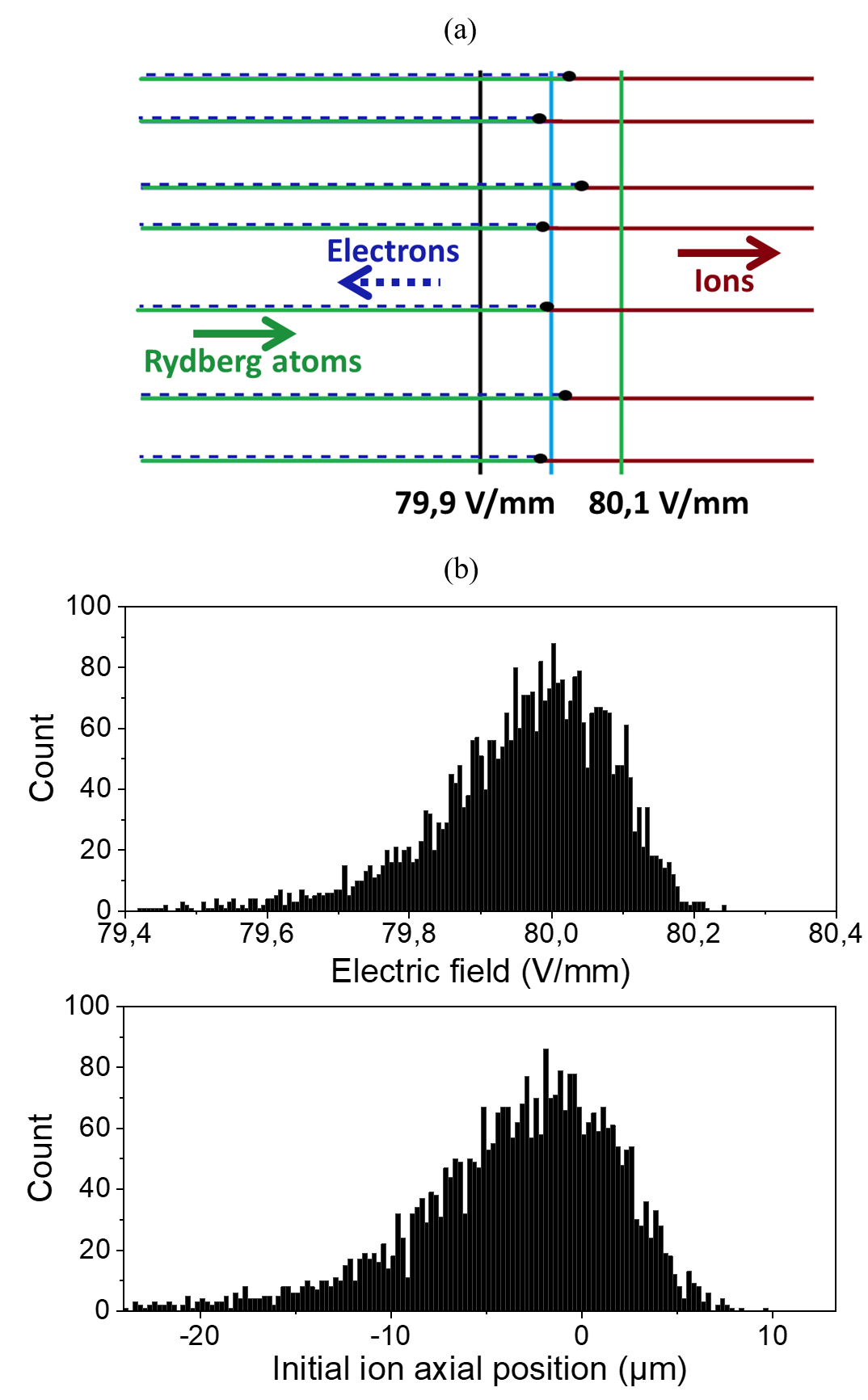}
    
    \caption{
        Simulation of the ionization process. (a) The figure illustrates representative particle trajectories before and after ionization. Rydberg atoms are shown in green, ions in red, and electrons in dashed blue. 
        (b) Initial ion distribution as a function of the electric field (top) and axial ionization position (bottom). The resonant ionization electric field is set to $F_{\rm ion} = 80.0~\mathrm V/mm$, corresponding to the relative position $x=0$. The resulting energy spread FWHM of the ion beam is approximately $1.0~\mathrm eV$.
    }
    \label{Ionization}
\end{figure}

To demonstrate the functionality of our code, Figures \ref{excitation} and \ref{Ionization} present simulation results where the initial atomic beam is modeled as a circularly distributed ensemble in a plane perpendicular to the propagation direction, with a radius slightly larger than the laser waist to optimize simulation time. Here, we simulate an ideal atomic beam of 10,000 cesium particles, without an initial transverse or longitudinal velocity spread ($V_{\text{transversal}} = 0\,\text{m/s}$ and $V_{\text{longitudinal}} = 250\,\text{m/s}$). These conditions allow us to isolate the effects of the excitation and ionization processes, free from the complexities of beam velocity distributions. The laser positions (located $-1\,\text{mm}$ from the longitudinal position of the resonant ionization electric field), waists ($W_0 = 50~\mu \text{m}$ for both lasers) and powers ($2\, \mu \text{W}$ for the $852\,\text{nm}$ laser and $200\,\text{mW}$ for the $512\,\text{nm}$ laser), along with the resonant electric fields for excitation ($F_0 = 71.056\,\text{V/mm}$, with $\alpha=2 \pi \cdot 10^{9}~\mathrm mm \, s^{-1}V^{-1}$) and ionization ($F_{\rm ion} = 80.0~\mathrm V/mm$, with a gradient of $2~\mathrm V/mm^2$), are configured to match realistic experimental conditions. In the program, the $x$-direction corresponds to the beam propagation axis, while the $y$ and $z$-directions represent the transverse components.

Figure~\ref{excitation} illustrates the excitation process in the presence of an electric-field gradient of $2~\mathrm{V/mm^2}$. Panel (a) shows that neutral atoms are promoted to a Rydberg state within this region, thereby optimizing the number of atoms reaching the Rydberg level and maximizing the resulting ion flux. In this three-level model, with spatially separated excitation and ionization zones, at most $33\%$ of the initial atomic population can be transferred to the Rydberg state. For the parameters used here, we achieve an excitation efficiency of only $\sim 17\%$, limited by the local field gradient and the available laser parameters (waists and power). Panel (b) displays the distribution of Rydberg excitation events as a function of electric field and axial position. Excitation occurs in a region where the field varies by only $0.005~\mathrm{V/mm}$ over the $2.5~\mu\mathrm{m}$-wide FWHM excitation zone, i.e., over a range smaller than the laser waist $W_0$. The combined effect of the controlled field gradient and the chosen Rydberg-state properties thus leads to a well-defined and reproducible excitation region. This approach also provides a convenient framework to study excitation efficiency for different Rydberg states, including high-$n$ states approaching the ionization continuum.

Figure~\ref{Ionization} illustrates the next stage of the simulation: the ionization process. The ionization probability increases as the local electric field experienced by the particles approaches the ionization threshold, $F_{\rm ion}$, as shown in Figure~\ref{ionization_rate}. Ionization occurs when particles enter the resonant ionization electric field region. In our Lua program, the ionization rate function, $\Gamma_{\rm ion}$, is given by Eq.~\eqref{gamma_ion_exp}, with $\Gamma_0 = 2\pi \cdot 8 \cdot  10^6~\mathrm s^{-1}$ and $\sigma = 0.1~\mathrm V/mm$. Panel (a) of Figure~\ref{Ionization} shows a zoomed-in view of the ionization zone, highlighting representative trajectories of ions (red) and electrons (dashed blue). Ions propagate to the right, while electrons move to the left. A spatial dispersion of ionization events is observed around the resonant field, $F_{\rm ion} = 80.0~\mathrm V/mm$. Panel (b) presents the initial ion distribution as a function of the electric field (top) and axial ionization position (bottom), showing that the ionization region spans an FWHM distribution of approximately $\sim 10~\mu \mathrm m$.

The distributions of excitation/ionization events and the resulting ion energy spectra are generally asymmetric. This asymmetry is expected because the ionization process is directional (atoms enter the ionization region from the low-field side) and because the field-ionization rate rises exponentially with the local field (see Eq.~\eqref{gamma_ion_exp}). The corresponding survival probability therefore suppresses early events and produces a longer tail on the high-field side. Since the extracted kinetic energy is a monotonic function of the ionization position in the extraction potential, the skewness of the ionization-position distribution is directly transferred to the measured and simulated kinetic-energy distributions. Formally, the event-position distribution is proportional to \(\Gamma_{\rm ion}(F(x))\exp\!\left[-\int\Gamma_{\rm ion}(F(x'))\,\frac{dx'}{v_x}\right]\), which directly yields a skewed distribution when \(\Gamma_{\rm ion}\) grows rapidly with \(F\). Here \(v_x\) denotes the axial velocity component along the propagation direction.

As demonstrated in Figures \ref{excitation} and \ref{Ionization}, the simulation provides detailed insights into excitation and ionization events --- information that is not directly accessible from experimental measurements. This enhanced level of detail significantly improves our understanding of particle dynamics and ion beam formation. The zoomed-in views and distributions presented in these figures illustrate the excitation and ionization zones, revealing position and velocity dispersions that directly influence the initial ionization conditions and contribute to chromatic and geometric aberrations.

In summary, this program enables a systematic study of various experimental parameters, including applied voltages, atomic beam characteristics, and laser parameters (such as geometry, wavelength, waist, and intensity), thereby enhancing the interpretation of experimental results. By providing a quantitative understanding of the underlying physical processes, the simulation serves as a powerful tool for optimizing experimental configurations and improving ion beam performance. Furthermore, the output of this Lua-based program can be directly incorporated into SIMION simulations, supplying realistic initial distributions for ion/electron positions and velocities, which are essential for accurate ion/electron beam modeling and optimization.

In the following section, we present examples applied to a FIB system, with particular emphasis on the ionization process and its influence on energy beam dispersion. 

\section{Application: ColdFIB}

\begin{figure}[!h]
    \centering
    \includegraphics[width=0.8\linewidth]{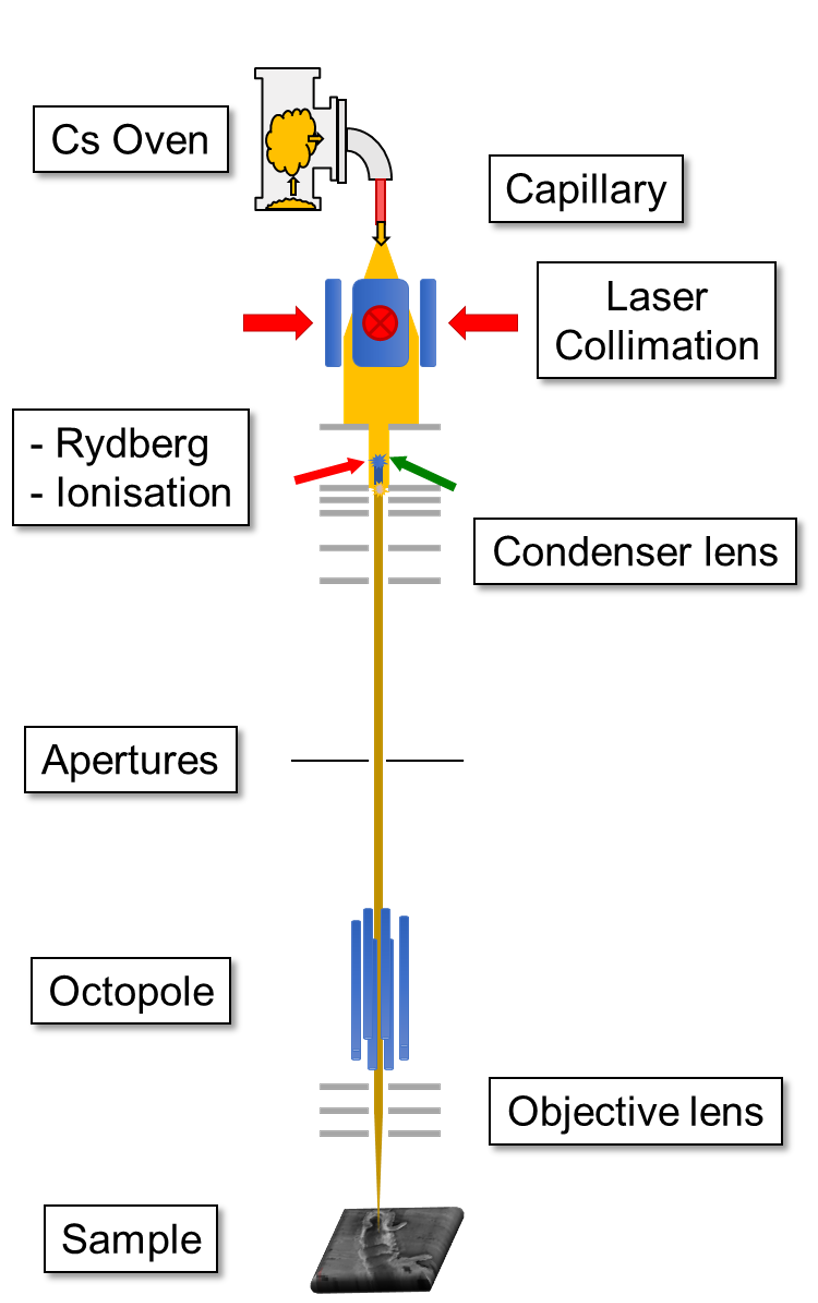}
    \caption{Schematic of the FIB setup used in the ColdFIB experiment, analyzed in this section through simulations. The ionization region is shown with more detail in Figure \ref{exemple_FIB_simion}.}
    \label{Coldfib}
\end{figure}

A key challenge in FIB applications is achieving nanometer-scale beam sizes while maintaining high ion flux, a task complicated by geometric and chromatic aberrations. Well-defined beam energy with minimal spread is crucial for generating monochromatic ion beams necessary for precise milling or imaging. As discussed, the initial axial position of the ions has a significant impact on energy dispersion and contributes to geometric aberrations. Minimizing the ionization region along the beam propagation axis reduces both the longitudinal energy spread and geometric distortions. The resulting ion distributions are highly sensitive to factors such as atomic velocity, the choice of Rydberg state, and the local electric field. The FIB system used in these simulations, ColdFIB (Figure \ref{Coldfib}), has been extensively characterized in previous studies \cite{viteau2016ion, antoni2018watt, kime2013high, reveillard2018coldfib}, and thus only a brief overview is presented here.

Cesium atoms from the thermal oven \cite{hahn2022comparative} are initially collimated by lasers along two transverse directions \cite{reveillard2022efficient} before being excited to Rydberg states via a two-photon excitation scheme, as illustrated in Figure \ref{exemple_FIB_simion}. The Rydberg atoms then propagate until they are ionized by the applied electric field, generating the ion beam. The resulting ions are extracted, accelerated, and focused using a two-lens system comprising a condenser lens and an objective lens. Between these lenses, double octupoles are employed to scan the sample and compensate for astigmatism. Additionally, apertures are used to select only the central region of the ion beam, thereby enhancing FIB resolution and image quality, though this comes at the cost of reduced beam current.

As discussed in the previous section, a key challenge in achieving a nanometer-scale beam in FIB systems is the presence of aberrations, particularly chromatic and geometric aberrations. To mitigate these effects, it is crucial to minimize both the initial energy and position dispersion of the ion beam, as well as its transverse temperature. This requires reducing the initial spread in the axial position and radial velocity of the ions. The simulated results for the final kinetic energy distribution presented in this work are taken at the aperture output (see Figure \ref{Coldfib}) as the particles in this area are already accelerated, so their energy remains effectively constant up to the sample.

The subsequent components of the FIB, such as the octupoles and objective lens, primarily serve to shape, orient, and focus the ion beam without significantly affecting its energy characteristics.

In the following, we investigate how the properties of the atomic source affect the excitation and ionization processes, which directly affect ion beam focusing, and present experimental images obtained with the ColdFIB setup. To this end, we consider and simulate two distinct types of initial atomic beams:
\begin{itemize}
    \item The effusive beam of the ColdFIB setup, using parameters based on current experimental data.
    \item A laser-cooled beam produced by a two-dimensional magneto-optical trap (2D-MOT), following the description in \cite{xie2022cold} and similar to the setup used in \cite{mitchell2025selective}.
\end{itemize}

 Next, using the ColdFIB effusive beam, we investigate the impact of selecting different intermediate Rydberg states during the ionization process by comparing two states, each characterized by a distinct ionization rate, $\Gamma_{\text{ion}}$. These rates produce two distinct sets of parameters for the exponential fit described in the previous section, which define the ionization rates used in our simulations.

 \begin{figure}[!h]
     \centering
     \includegraphics[width=1\linewidth]{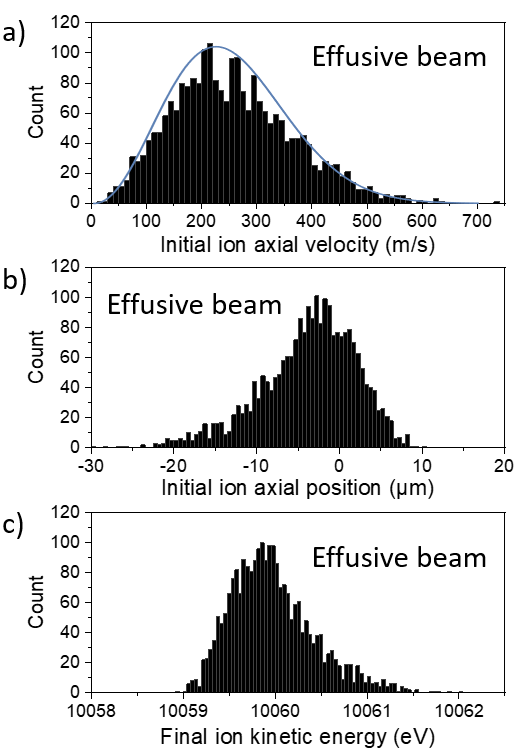}
     \caption{The atomic beam is modeled using a Maxwell-Boltzmann velocity distribution defined by the most probable velocity, $v_\text{mp} = 238~\text{m/s}$. Panel (a) shows the initial axial velocity distribution of the ions for the cesium effusive beam, along with the fitted Maxwell-Boltzmann curve (blue) corresponding to the atomic velocity distribution. Panels (b) and (c) display the initial longitudinal position dispersion and the resulting final kinetic energy of the ions, respectively. The bin sizes for these distributions are $10~\mathrm m/s$ for panel (a), $0.5~\mu \mathrm m$ for panel (b), and $0.04~\mathrm eV$ for panel (c).}
     \label{effusif_position_velocity}
 \end{figure}

As shown in Figure \ref{effusif_position_velocity}(a), the velocity distribution of the effusive cesium beam follows a Maxwell-Boltzmann distribution for the velocity magnitude, with the most probable velocity $v_{\rm mp} = 238~\mathrm m/s$ for atoms emerging from an oven at $180^\circ\mathrm C$ with a nozzle at $300^\circ\mathrm C$. Although an effusive beam naturally obeys a Maxwell-Boltzmann distribution, SIMION does not natively support this. In our Lua program, we implement it using a rejection sampling method: candidate velocities are initially drawn from a uniform distribution between $0~\mathrm m/s$ and $800~\mathrm m/s$, the maximum of the Maxwell-Boltzmann distribution is evaluated at the most probable velocity, and a random number between $0$ and this maximum is generated for each candidate. Thus, each drawn velocity is accepted if the random number is below the normalized Maxwell-Boltzmann value at that velocity magnitude, otherwise it is rejected.

We model the initial positions by introducing an ideal point source at the origin from which all particles are emitted. The beam is assumed to have cylindrical symmetry along the $x$ axis and a uniform angular distribution within a given emission cone. The maximum angular spread is set by the geometry of the oven nozzle, its distance from the excitation region, and the laser waists, which together determine the range of emission angles. For each particle, we draw two random angles in spherical coordinates to define the velocity direction and assign a velocity magnitude from the Maxwell-Boltzmann distribution. Using the known distance between the point source and the laser plane, we then propagate the particle to an initial position located $3~\mathrm{mm}$ upstream the region where the excitation beams cross. As a result, the initial spatial distribution is determined not only by the oven temperature, but also by the experimental geometry (nozzle size, source-laser distance, and laser waists), under the assumption of cylindrical symmetry. This construction provides a consistent definition of the longitudinal velocity component along the beam axis ($x$) and of the transverse components ($y$ and $z$), all drawn from the Maxwell-Boltzmann distribution and constrained by the imposed symmetry.

Our primary interest lies in the initial longitudinal distribution of the ions, which is crucial for optimizing the FIB and for the study we aim to conduct. The laser positions, waists, powers, and the excitation and ionization resonant fields are configured to closely match the experimental conditions. The resulting initial ion velocity and position distributions are shown in Figure \ref{effusif_position_velocity}.

For this simulation, 14341 neutral particles were initially launched, following the conditions described above. The ionization rate function, $\Gamma_{\rm ion}$, is calculated using Eq.~\eqref{gamma_ion_exp}, with $\Gamma_0 = 2\pi \cdot 8 \cdot  10^6~\mathrm s^{-1}$, $F_{\rm ion} = 80.0~\mathrm V/mm$, and $\sigma = 0.1~\mathrm V/mm$. Under these parameters, approximately 2385 particles were ionized, yielding an ionization efficiency of 17\%. This efficiency is primarily determined by the Rydberg excitation probability, which depends on the 200~mW power of the green laser used ($\lambda = 512~\mathrm nm$). Because the excitation and ionization regions are spatially separated, only particles that reach the ionization zone in the Rydberg state --- particularly the fastest ones --- are effectively ionized by the electric field, since particles may undergo de-excitation during transit. As a result, the ionization efficiency is influenced not only by the Rydberg state but also by the $1~\mathrm mm$ spatial separation between the excitation and ionization zones.

The width of the initial axial position distribution of the resulting ions, shown in Figure~\ref{effusif_position_velocity}(b), is largely determined by the parameter $\sigma = 0.1~\mathrm V/mm$, which sets the slope of the exponential ionization rate function $\Gamma_{\rm ion}$ (Eq.\eqref{gamma_ion_exp}). In this simulation, the FWHM of the initial longitudinal ion distribution is $10~\mu$m, yielding a FWHM of the final kinetic energy distribution of $0.8~\mathrm eV$, as illustrated in Figures\ref{effusif_position_velocity}(b-c). These results closely reproduce the experimental data previously obtained with this setup \cite{viteau2016ion,reveillard2018coldfib,reveillard2022efficient}.

\begin{figure}[!h]
    \centering
    \includegraphics[width=1\linewidth]{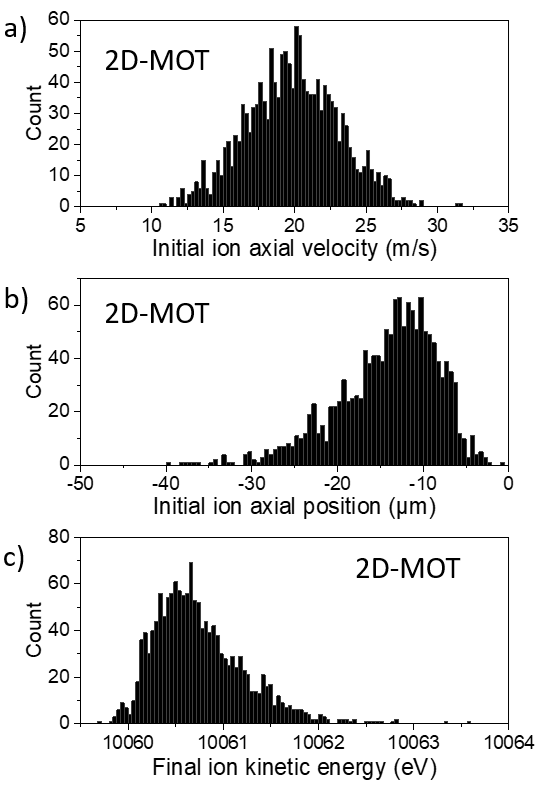}
    \caption{The atomic beam for the 2D-MOT is modeled using a Gaussian velocity distribution centered on the most probable velocity, $v_\text{mp} = 19.4~\text{m/s}$, with an FWHM of $8.1~\text{m/s}$, based on experimental values extracted from \cite{mitchell2025selective, xie2022cold, grimm2000optical}. Panel (a) shows the initial axial velocity dispersion of the ions. Panels (b) and (c) display the initial axial position dispersion and the resulting final kinetic energy of the ions, respectively. The bin sizes for these distributions are $0.25~\mathrm m/s$ for panel (a), $0.5~\mu\text{m}$ for panel (b), and $0.04~\mathrm eV$ for panel (c).}
    \label{MOT2D_separated}
\end{figure}

To benchmark the effusive beam against alternative sources, we now consider a standard cesium 2D-MOT operated under otherwise identical conditions (laser configurations, electric fields, etc.), and compare the results to those obtained with the effusive beam. For the 2D-MOT, the velocity magnitude is assumed to follow a Gaussian distribution centered at the most probable velocity, $v_\text{mp} = 19.4~\mathrm m/s$, with an FWHM of $8.1~\mathrm m/s$ (Figure \ref{MOT2D_separated}(a)). The 3D-projection of the velocity distribution is defined using the same method described for the effusive beam. These parameters are consistent with experimental measurements reported in \cite{mitchell2025selective, xie2022cold, grimm2000optical}. 

For this simulation, 14341 neutral particles were launched using the specified initial conditions. Approximately 1477 particles were ionized, corresponding to an ionization efficiency of 10\%. In the case of the 2D-MOT, compared to the effusive beam, a larger fraction of Rydberg-excited atoms undergoes spontaneous de-excitation before reaching the ionization region, which reduces the overall ionization efficiency. Despite this limitation, the FWHM of the initial axial position distribution of the produced ions, shown in Figure \ref{MOT2D_separated}(b), remains approximately $10~\mu\rm m$, yielding a FWHM of the final ion kinetic energy distribution of $0.8~\mathrm eV$, as shown in Figure \ref{MOT2D_separated}(c). These results are similar to those obtained with the effusive beam, except for the reduced ionization efficiency, which consequently lowers the final ion flux---a critical parameter for standard FIB applications.

A first approach to mitigate the loss of ion flux is to accelerate the 2D-MOT atoms while preserving the initial transverse and longitudinal characteristics of the atomic beam. Alternatively, the accelerated 2D-MOT source can be replaced by a supersonic beam of cesium atoms, transversely cooled using the same laser configuration as in our ColdFIB setup. Both alternatives were simulated using the same parameters as previously described. 

For the accelerated 2D-MOT, the mean velocity magnitude is set to $v_\text{mp} = 250~\mathrm m/s$ with an FWHM of $5~\mathrm m/s$. For the supersonic beam, the mean velocity is $v_\text{mp} = 600~\mathrm m/s$ with an FWHM of $60~\mathrm m/s$. The initial velocity dispersions of these beams are modeled using Gaussian distributions.

The simulation results are consistent with previous findings: the FWHM of the initial axial position distribution of the ions is approximately $10~\mu\rm m$, leading to a final kinetic energy distribution with an FWHM of approximately $0.8~\mathrm eV$ in both cases. However, the ionization efficiencies differ significantly. For the accelerated 2D-MOT, about $27\%$ of the neutral atoms are ionized, while for the supersonic beam, only $12\%$ are ionized under the same parameters. This reduction in efficiency arises because higher longitudinal velocities reduce the excitation efficiency to Rydberg states. To compensate for this, increasing the green laser power (e.g., from $200~\mathrm mW$ to $4~\mathrm W$) boosts the ionization efficiency of the supersonic beam to approximately $26\%$.

These results confirm that while increasing the longitudinal velocity of the atomic beam --- either through acceleration in the 2D-MOT or by using a supersonic beam --- does not significantly affect the spatial or energy spread of the produced ions, it strongly influences the ionization efficiency. Optimizing laser power and Rydberg excitation parameters is therefore crucial to maintain high ion flux while preserving narrow distributions in position and energy, both of which are critical for FIB applications. 

A second strategy to mitigate the loss of ion flux in a 2D-MOT configuration is to reduce the spontaneous de-excitation of Rydberg atoms, either by decreasing the distance between the excitation and ionization regions or by increasing the principal quantum number $n$ of the selected Rydberg state, thereby extending its lifetime. In practice, however, increasing $n$ tends to reduce the excitation efficiency, which limits the effectiveness of this approach. We therefore choose to minimize the separation between the excitation and ionization zones while keeping all other parameters fixed. In this configuration, the two regions are effectively superposed, so that atoms, once excited, immediately encounter the ionizing electric field. Experimentally and in the simulations, this is realized by tuning the wavelength of the green laser such that the resonant electric field for excitation coincides with that for ionization. This scheme ensures a nearly continuous transition from excitation to ionization, strongly suppressing intermediate de-excitation events and, in principle, allowing ionization efficiencies of about $100\%$ for atoms that reach the Rydberg state. By contrast, when the regions are spatially separated, the maximum achievable ionization rate is limited by  the excitation process to $\sim 33\%$ in our three-level model, and with the parameters used here we obtain only $\sim 10\%$, constrained by the local field gradient and the available laser waists and powers.

\begin{figure}[!ht]
    \centering
    \includegraphics[width=1\linewidth]{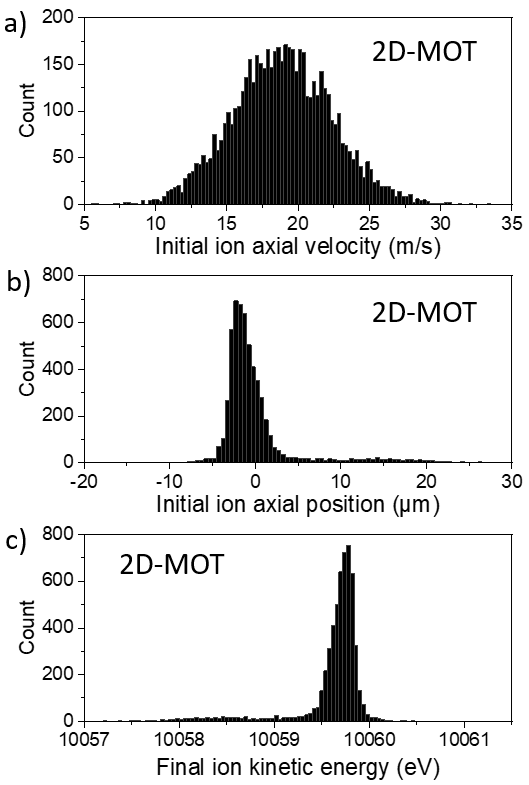}
    \caption{The beam parameters for the 2D-MOT are identical to those described in Figure \ref{MOT2D_separated}, except that the excitation and ionization zones are superposed to maximize the ionization efficiency. Panel (a) shows the initial axial velocity distribution of the ions. Panels (b) and (c) display the initial axial position distribution and the resulting final kinetic energy distribution of the ions, respectively. The bin sizes for these distributions are $0.25~\mathrm m/s$ for panel (a), $0.5~\mu\text{m}$ for panel (b), and $0.04~\mathrm eV$ for panel (c).}
    \label{MOT2D_overlapping}
\end{figure}

To simulate this configuration, 14341 neutral particles were launched using the same procedure and initial conditions as for the 2D-MOT with separated excitation and ionization zones. As shown in Figure \ref{MOT2D_overlapping}(a), the longitudinal velocity distributions are identical to those used previously (Figure \ref{MOT2D_separated}(a)). Approximately 5697 particles were ionized under this configuration, corresponding to an ionization efficiency of 40\%, nearly four times higher than the 10\% efficiency obtained with separated excitation and ionization zones. This ionization rate is limited by the Rydberg excitation efficiency, which is lower than in the separated case due to the steeper electric-field gradient reducing the probability of excitation to the Rydberg state. Adding to this the same experimental constraints described before (laser powers and waists), the idealized $100\%$ ionization efficiency cannot therefore be reached.

Nevertheless, these results demonstrate that overlapping the excitation and ionization zones effectively increases the ion flux for a 2D-MOT, providing a practical alternative to the technical complexity and cost associated with an accelerated 2D-MOT or a supersonic atomic beam.

An additional key advantage revealed by these simulations is the substantial reduction of the FWHM of both the initial axial position and the final kinetic energy distributions of the produced ions. As shown in Figure \ref{MOT2D_overlapping}(b), the FWHM of the initial ion axial position distribution is approximately $3.5~\mu\text{m}$, resulting in a FWHM of the final ion kinetic energy distribution of $0.3~\text{eV}$ (Figure \ref{MOT2D_overlapping}(c)). Compared to the 2D-MOT configuration with separated excitation and ionization zones, this represents almost a threefold reduction in longitudinal dispersions. This improvement enhances FIB performance by increasing ionization efficiency and minimizing chromatic aberrations, thereby enabling tighter beam focusing and improved spatial resolution.

The primary factor behind this improvement is the superposition of the excitation and ionization regions, which ensures that excited neutral particles are immediately exposed to the ionizing electric field. In contrast, in the configuration with separated zones, atoms remaining in the Rydberg state are more likely to ionize slightly before reaching the ionization threshold, resulting in a broader ion distribution along the propagation axis. By overlapping the excitation and ionization regions, the excited Rydberg atoms are generated near the resonant electric field. Due to the steep electric field gradient, ionization occurs within a narrow spatial region, resulting in a narrower distribution in both initial position and final kinetic energy, ultimately leading to a more controlled and focused ion beam.

This outcome prompts the question of whether similar results, particularly the reduction in longitudinal dispersions of both initial position and final kinetic energy of the ions, can be achieved with an effusive beam. To explore this, we conducted a similar simulation using an effusive beam as the source of neutral cesium atoms, while superimposing both the excitation and ionization regions. In this configuration, once the atoms are excited, the Rydberg atoms immediately encounter the electric field required for ionization, eliminating any intermediate de-excitation.

\begin{figure}[!h]
    \centering
    \includegraphics[width=1\linewidth]{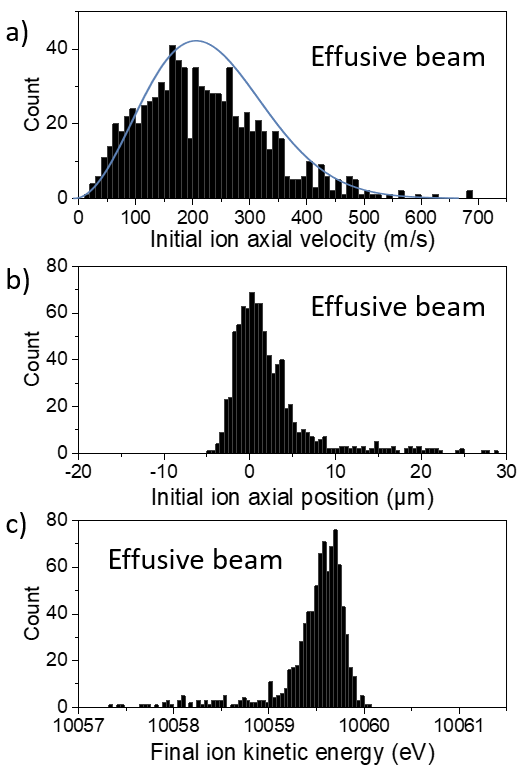}
    \caption{The beam parameters are identical to those described in Figure \ref{effusif_position_velocity}, except that the excitation and ionization zones are superposed. Panel (a) shows the initial axial velocity distribution of the ions for the cesium effusive beam, along with the fitted Maxwell-Boltzmann curve (blue) corresponding to the atomic velocity distribution (see Figure \ref{effusif_position_velocity}). Panels (b) and (c) display the initial axial position distribution and the resulting final kinetic energy distribution of the ions, respectively. The bin sizes for these distributions are $10~\mathrm m/s$ for panel (a), $0.5~\mu\text{m}$ for panel (b), and $0.04~\mathrm eV$ for panel (c).}
    \label{EffusiveBeam_overlapping}
\end{figure}

For this simulation, 14341 neutral particles were launched, following the same procedure and initial conditions as in the previous simulation. However, as shown in Figure \ref{EffusiveBeam_overlapping}(a), the shape of the longitudinal initial velocity distribution in this configuration differs from that of the effusive beam in the separated configuration (see Figure \ref{effusif_position_velocity}(a)). In this setup, approximately 867 particles were ionized, corresponding to an ionization efficiency of 6\%, which is nearly three times lower than the 17\% efficiency observed when the excitation and ionization zones are separated. 

These results demonstrate that, for the set of parameters used, superposition of the excitation and ionization regions is less effective, resulting in a substantial reduction in the ion flux. However a notable outcome of these simulations is the substantial reduction in the FWHM of both the initial axial position and the final kinetic energy distributions of the produced ions. As shown in Figure \ref{EffusiveBeam_overlapping}(b), the FWHM of the initial ion axial position distribution in this configuration is approximately $6~\mu\text{m}$, leading to a FWHM of $0.5~\text{eV}$ for the final ion kinetic energy distribution, as illustrated in Figure \ref{EffusiveBeam_overlapping}(c). These results are significant because, for the given set of parameters, the longitudinal dispersions in both initial position and final kinetic energy are reduced by a factor of two compared to the configuration with separated excitation and ionization zones. This reduction markedly diminishes chromatic aberrations, improving the FIB resolution. 

The ionization efficiency in the effusive beam configuration differs from that in the 2D-MOT due to the broader velocity distribution of the effusive beam. In the superposed configuration, Rydberg atoms are generated near the resonant electric field, which enhances the ionization probability for slower particles, as they interact with the lasers for a longer duration. As shown in Figure \ref{EffusiveBeam_overlapping}, the steep electric field gradient ensures that most slower Rydberg atoms are ionized within a narrow spatial region, resulting in a reduced final kinetic energy spread.

Now that we have examined the influence of the atomic source velocity characteristics on the initial ion positions and final kinetic energy distributions, we turn to the effect of the chosen Rydberg state during the ionization process. For this second test, we simulate, in the separated zones configuration, two distinct Rydberg states, each characterized by a different ionization rate $\Gamma_{\text{ion}}$, while keeping the effusive beam parameters unchanged (as described in Figure \ref{effusif_position_velocity}). The ionization rates were computed using the exponential fit described in Eq.~\ref{gamma_ion_exp}, which depends on the following key parameters: $\Gamma_{0}$, the ionization rate of a given Rydberg state near the ionization threshold electric field $F_{\rm ion}$, and $\sigma$, the slope of the ionization rate in this region. 

 \begin{figure}[!h]
    \centering
    \includegraphics[width=1\linewidth]{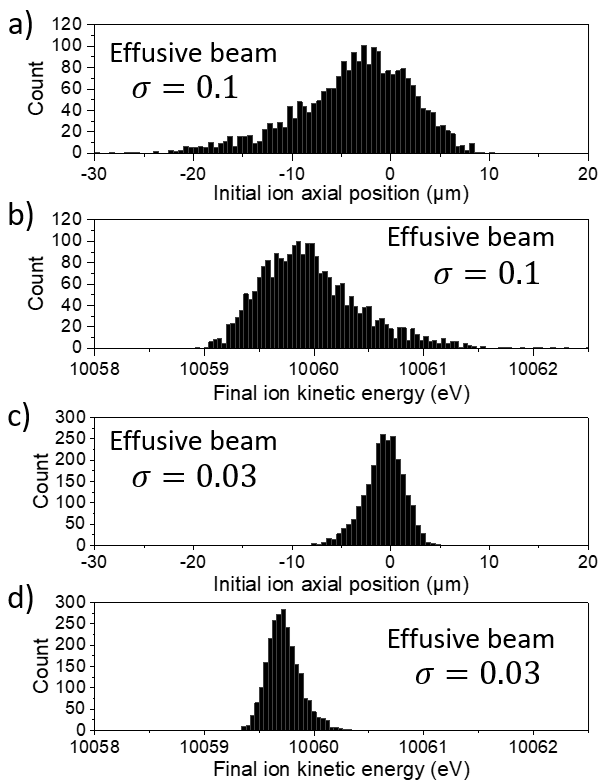}
    \caption{Initial axial position and final kinetic energy distributions of ions for two different Rydberg states, each characterized by a distinct slope $\sigma$ of the ionization rate in the separated-zones configuration. Panels (a,b) (same conditions as in Figure~\ref{effusif_position_velocity}(b,c)) correspond to an ionization rate $\sigma = 0.1~\mathrm{V/mm}$, yielding an ionization region with an FWHM of $\sim 10~\mu\mathrm{m}$ in the initial axial position and a final kinetic-energy spread with an FWHM of $\sim 0.8~\mathrm{eV}$. Panels (c,d) correspond to $\sigma = 0.03~\mathrm{V/mm}$, which produces an ionization region with an FWHM of $\sim 4~\mu\mathrm{m}$ and a kinetic-energy FWHM of $\sim 0.3~\mathrm{eV}$. These distributions illustrate how the slope of the ionization rate directly controls the spatial localization of ionization events and the resulting energy spread. The bin sizes are $0.5~\mu\mathrm{m}$ for panels (a,c) and $0.04~\mathrm{eV}$ for panels (b,d).}
    \label{sigma_coupling}
\end{figure}

In practice, varying $\sigma$ corresponds to selecting different Rydberg states. As shown in Figure \ref{ionization_rate} (examples A and B), choosing a different ionizing Rydberg state generally requires small adjustments to both the laser wavelength and the resonant ionizing electric field, which modify the local Rydberg Stark slope in the ionization region. This slope is critical because it determines the spatial extent of the ionization zone: a steeper slope results in a more localized ionization region.

For both coupling scenarios, we employed the same effusive beam parameters described previously, where the excitation and ionization zones are separated. Two values of the coupling parameter, $\sigma = 0.1~\mathrm V/mm$ and $\sigma = 0.03~\mathrm V/mm$, were simulated, and the resulting distributions of the initial axial position and final kinetic energy of the ions are shown in Figure \ref{sigma_coupling}. In the weaker coupling case ($\sigma = 0.1~\mathrm V/mm$), the ionization region exhibited a spatial dispersion of approximately $10~\mu\rm m$, consistent with the results in Figure \ref{effusif_position_velocity}. In contrast, for $\sigma = 0.03~\mathrm V/mm$, corresponding to a steeper local Rydberg Stark slope, the ionization region became narrower, with an FWHM of $4~\mu\rm m$. The final kinetic energy dispersion decreased from a FWHM of $0.8~\mathrm eV$ to $0.3~\mathrm eV$ between the two cases. These results indicate that stronger coupling, i.e., a steeper local Rydberg Stark slope in the ionization zone, significantly improves the spatial localization of ionization events and reduces the final energy spread of the ion beam. However, this configuration can also limit the ionization efficiency, as the Rydberg state lifetime decreases under these conditions. 

   \begin{figure}[!h]
    \centering
    \includegraphics[width=1\linewidth]{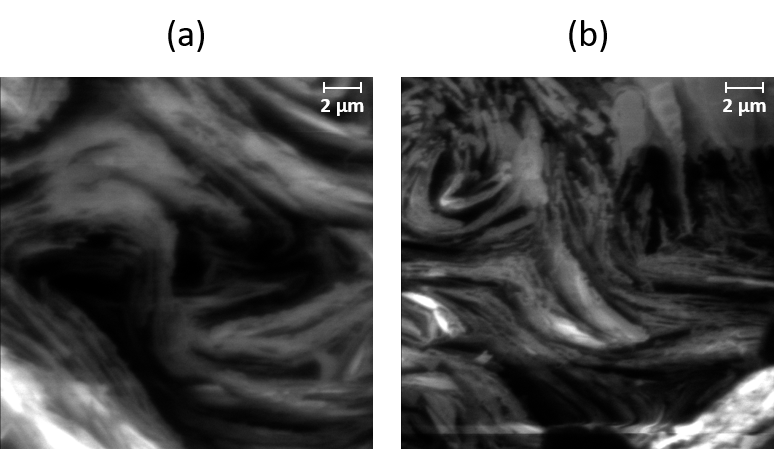}
    \caption{
    Experimental secondary-electron images of the same graphite sample acquired with the ColdFIB instrument under identical column settings (ion energy $10~\mathrm{keV}$, same apertures and scan conditions), with focusing and astigmatism corrections optimized using the same procedure in both cases. The only change is the selected Rydberg state used for field ionization. (a) Rydberg state~A, measured kinetic-energy dispersion $\Delta E_{\rm FWHM}=0.8~\mathrm{eV}$. (b) Rydberg state~B, $\Delta E_{\rm FWHM}=0.3~\mathrm{eV}$. The improved sharpness in (b) is consistent with the simulations (Fig.~\ref{sigma_coupling}), which predict a more localized ionization region (smaller $\Delta x_{\rm ion}$). Energy spreads are obtained by retarding-field analysis (RFA). Field of view: $20~\mu\mathrm m$ (scale bar shown).
    }
    \label{images_stark}
 \end{figure}

The simulations predict that selecting a Rydberg state with a steeper effective ionization slope (smaller \(\sigma\)) localizes ionization events, reducing \(\Delta x_{\rm ion}\) and therefore \(\Delta E\). In addition, because ions are created within the extraction/first-lens field, a reduced and shifted creation region also mitigates geometric aberrations associated with non-paraxial extraction fields. The improved image sharpness in Fig.~\ref{images_stark} is therefore consistent with the combined reduction of chromatic aberrations (through smaller \(\Delta E\)) and geometric aberrations (through a more localized creation volume).

The images were obtained with the ColdFIB setup, where a graphite sample is probed using ions produced from two different Rydberg states. A focused ion beam is raster-scanned over the surface; the detected signal is the secondary-electron yield induced by ion impact and collected by the Everhart--Thornley detector. In the ColdFIB system, a tightly focused ion beam is raster-scanned across the sample surface using beam-steering optics (double-octopole system in Figure~\ref{Coldfib}). Secondary electrons emitted upon ion impact are collected with an Everhart--Thornley detector at $45^{\circ}$. The dwell time per pixel is tunable between $100~\mathrm ns$ and $3.2~\mathrm ms$, enabling control of the delivered dose and the signal-to-noise ratio. The system operates under high vacuum (approximately $8\times10^{-7}~\mathrm mbar$ in the oven region and $10^{-7}~\mathrm mbar$ elsewhere) to minimize scattering.

 The images in Figure~\ref{images_stark} clearly reveal an enhanced focusing performance, indicative of a smaller spot size and reduced aberrations. This improvement is corroborated by the measured dispersions of the final ion kinetic energies, obtained via retarding-field analysis (RFA), which agree with the simulated FWHM values of $0.8~\mathrm{eV}$ for the left image and $0.3~\mathrm{eV}$ for the right image. These results underscore the accuracy and versatility of our Lua-based simulation code: by generating realistic initial distributions of position, velocity, and energy associated with Rydberg ionization, it provides physically grounded input for \textsc{SIMION}, enabling precise trajectory calculations for the resulting charged particles and quantitative optimization of the experimental configuration.

\begin{table*}[t]
\caption{
Summary of the source configurations, simulations, and key outcomes.
}
\label{tab:summary_cases}

\renewcommand{\arraystretch}{1.18}
\begin{tabular*}{\linewidth}{@{\extracolsep{\fill}}lllllll}
\hline\hline
\textbf{Source} &
\textbf{Zones} &
\textbf{Speed Distribution} &
\textbf{Ion Fraction} &
\textbf{$\sigma$ (\si{V/mm})} &
\textbf{$\Delta x_{\rm ion,FWHM}$} &
\textbf{$\Delta E_{\rm FWHM}$} \\
\hline
Effusive & Separated  & MB, $v_{\rm mp}=\SI{238}{m/s}$ & $2385/14341$ (17\%) & 0.10 & $\SI{10}{\micro m}$  & $\SI{0.8}{eV}$ \\
2D-MOT   & Separated  & Gauss., $v_{\rm mp}=\SI{19.4}{m/s}$ & $1477/14341$ (10\%) & 0.10 & $\SI{10}{\micro m}$  & $\SI{0.8}{eV}$ \\
2D-MOT   & Overlapped & Gauss., $v_{\rm mp}=\SI{19.4}{m/s}$ & $5697/14341$ (40\%) & 0.10 & $\SI{3.5}{\micro m}$ & $\SI{0.3}{eV}$ \\
Effusive & Overlapped & MB, $v_{\rm mp}=\SI{238}{m/s}$ & $867/14341$ (6\%)   & 0.10 & $\SI{6}{\micro m}$   & $\SI{0.5}{eV}$ \\
Effusive & Separated  & MB, $v_{\rm mp}=\SI{238}{m/s}$ & $2385/14341$ (17\%) & 0.03 & $\SI{4}{\micro m}$   & $\SI{0.3}{eV}$ \\
\hline\hline
\end{tabular*}

\vspace{0.4ex}
\footnotesize\noindent
\emph{Notes:} ``Zones'' indicates whether excitation and field-ionization regions are spatially separated or overlapped.
Field ionization is modeled as $\Gamma_{\rm ion}(F)=\Gamma_0\exp[(F-F_{\rm ion})/\sigma]$ [Eq.~(\ref{gamma_ion_exp})], where $\sigma$ sets the characteristic field scale for an $e$-fold change of $\Gamma_{\rm ion}$ and therefore the axial extent of the ionization region for a given field gradient.
The extracted spread $\Delta E_{\rm FWHM}$ is primarily set by $\Delta x_{\rm ion,FWHM}$ through $\Delta E\simeq F_{\rm ion}\Delta x_{\rm ion}$ (in eV unit).
MB = Maxwell--Boltzmann; Gauss. = Gaussian.
\end{table*}

\section{Conclusion and Perspectives}

In this study, we have investigated the excitation and ionization dynamics of cesium atoms using a custom Lua program integrated within the SIMION simulation platform. This hybrid approach enables the simultaneous modeling of internal quantum-state dynamics and external particle trajectories, providing a comprehensive analysis of the ionization process. By simulating various initial atomic beams, including effusive sources and laser-cooled beams, we demonstrated how precise control of experimental parameters directly influences ion beam characteristics. Key optimizations achieved include reduced energy dispersion, minimized axial position spread, and higher ionization efficiency, all of which were validated by experimental results obtained from the ColdFIB setup. Tab.~\ref{tab:summary_cases} presents a summary of these simulations.

The Lua-SIMION framework has proven to be both versatile and predictive, accurately reproducing experimental observations across a variety of configurations. By coupling internal quantum transitions with external field interactions in a single simulation, this approach enables advanced studies extending beyond traditional FIBs. Looking forward, the ability to combine theoretical insights with experimental observations provides a robust foundation for advancing both FIB technology and precision quantum measurements. 

The Lua-SIMION platform is a powerful tool for advancing both fundamental research and applied technological challenges in atomic, molecular, and optical physics. It offers new possibilities for applications such as ion-electron correlation analysis in dual-microscopy, where real-time aberration corrections and enhanced resolution are crucial \cite{lopez2019real}. The platform is also suitable for simulating charged-particle dynamics in Paul traps, molecular beam experiments, ultracold plasma dynamics, and the study of novel ionization mechanisms driven by tailored laser fields.

\section*{Code availability}
The Lua package developed in this work is available at \url{https://doi.org/10.5281/zenodo.19065668} or \url{https://src.koda.cnrs.fr/daniel.comparat.1/simion-lua-article-2026}. The repository includes the excitation/ionization algorithms, example input files, and a demonstrative \textsc{SIMION} geometry reproducing the workflows and figures reported here. Simulations were performed with \textsc{SIMION}~8.2 (2020). The full ColdFIB column CAD is not distributed due to engineering/IP constraints; the public demonstrator preserves the relevant field-gradient features that govern excitation and field ionization. The code is released under the \texttt{MIT} No Attribution License.

\section{Acknowledgements}

This work was supported by funding from the Agence Nationale de la Recherche (ANR) under project No. ANR-21-CE42-0010-01 (FIBBack) and the European Research Council (ERC) under grant agreement No. CITRON 101055250.

\appendix 
\section{Master equation}
\label{KMC_algo}
The goal is to solve the master equation
\begin{equation}
    \frac{dP_k}{dt} =  \sum_{l=1}^N \Gamma_{k\leftarrow l} P_l - \sum_{l=1}^N  \Gamma_{l\leftarrow k}  P_k 
    \label{mastereq_app}
\end{equation}
This equation can be written as a matrix form $\frac{d \mathbf{P} }{dt}  = - \mathbf{\Gamma}  \mathbf{P}$.
It is beyond the scope of this article to review all interpretations, methods, and algorithms based on the master equation (see  \cite{fichthorn1991theoretical,binder2019monte,landau2021guide} for more details), but in our case, we will interpret it as describing the time evolution of the probability $P_k$ of a system to occupy each one of a discrete set of states numbered by $k$.  Each process occurs at a certain rate $\Gamma_{l\leftarrow k} (t)$. In our case, we have a physical system with a given particle, at time $t$, in a given state $|k\rangle$, and we would like to follow its evolution, that is, to know if it evolves and if it does, when, and toward which state $|l\rangle$.

From this point of view, one of the simplest algorithms to solve this equation is the Monte Carlo stochastic sampling fixed time-step algorithm based on explicit (or forward) Euler discretization of the master equation:
$ P_k(t+\Delta t) \approx P_k(t) + \sum_{l=1}^N \Gamma_{k\leftarrow l}(t)  P_l (t) \Delta t - \sum_{l=1}^N  \Gamma_{l\leftarrow k}(t) P_k(t) \Delta t    $. Thus, for the system being in state $|k\rangle$ at time $t$, the choice to evolve at time $t+\Delta t$ is made by a Metropolis type of algorithm, that is, the state evolves only if $ \Gamma_{k}(t) \Delta t \ge u $ where  $u$ is taken randomly from a uniform unit interval $0<u\leq 1$ 
and  $\Gamma_{k } = \sum_l \Gamma_{l\leftarrow k}$ is the total evolution rate of state $|k\rangle$.
The main disadvantage of this method \cite{ceccatto1986effective,jansen2003introduction} is that  $\Delta t$ has to be small enough to maintain accuracy, meaning  $\Gamma_{k}(t)  \Delta t \ll 1$. 
This is easily illustrated by the case of a constant $\Gamma_{k}$, for which the survival function $S(t) = e^{-\Gamma_{k} t}$, that is the probability that the system remains unchanged (survives) up to $t$, is approximate by the Euler's (Poisson process) method by $S_{\rm Euler}(t=n \Delta t) = (1- \Gamma_{k} \Delta t)^n = (1- \Gamma_{k} \Delta t)^{t/\Delta t}$, that converge to $S(t)$ only if $\Gamma_{k} \Delta t \ll 1$.

Outside of static cases, a common solution to this problem is thus to move to the Kinetic Monte Carlo  (KMC)  algorithm -- sometimes called  Bortz-Kalos-Lebowitz \cite{bortz1975new}, Gillespie  stochastic simulation
algorithm (SSA)  \cite{gillespie1977exact}, n-fold way or dynamic Monte Carlo --
by choosing the  time step $t'$, during which the evolution of the system occurs and is calculated  by solving
	$ \exp \left( - \int_t^{t'} \Gamma_{k }(\tau) d \tau  \right)  =  u$ where $u$ is taken randomly from  an uniform unit-interval $0<u\leq 1$.
    Several improvements of the KMC methods have been proposed, called First Reaction Method (FRM), Next Reaction Method (NRM), Random Selection Method (RSM),
optimized direct method (ODM), sorting direct method
(SDM), composition-rejection SSA, ... \cite{prados1997dynamical,anderson2007modified,thanh2015simulation,andersen2019practical,chittari2024revisiting}. 
The key expression is the reaction time formula that comes from the fact that
for a
system initially at time $t$ in state $|k\rangle$,
the probability that the system
has not yet escaped from state $|k\rangle$ at time $t'$ is given by
$ \exp \left( - \int_t^{t'} \Gamma_{k }(\tau) d \tau  \right) $.
At time $t'$, a reaction takes place, and a new state $|l\rangle$ can be generated by picking it out of all possible new states with a probability proportional to $\Gamma_{l\leftarrow k} (t')$. 
The state evolution from $|k\rangle$ to $|l\rangle$ is justified due to the Markovian behavior of the master equation, so at time just before $t'$, the system is still in state $|k\rangle$ and its evolution rates are given by $\Gamma_{l\leftarrow k} (t')$. The KMC method makes exact numerical calculations and is indistinguishable from the behavior of the real system, reproducing for instance, all possible data in an experiment, including its statistical noise. The gain compared to the Euler's method is clearly illustrated by the case of a constant $\Gamma_{k}$, where a single step is enough to find the evolution time 
$ t'    =  t - (\ln u)/ \Gamma_{k}$ 
in the KMC method whereas it takes $n\approx (t'-t)/\Delta t \sim 1/(\Gamma_{ k} \Delta t) \gg 1 $ steps in the Euler's method.

However, for time varying $\Gamma_{ k}(t)$, it is difficult to calculate or estimate the time integration  $\int_t^{t'}  \Gamma_{ k}(\tau) d \tau$. For instance in Ref. \cite{prados1997dynamical}, the method used was the rectangle rule with fixed time steps to cut the integral into small parts, each estimated using $ \int_t^{t+\Delta t} \Gamma_{ k}(\tau)\, d\tau \approx \Delta t\Gamma_{ k}(t) $.

The error for the timing $\Delta t$ can be estimated based on the Taylor-Lagrange's expansion
\begin{align*}
\int_t^{t+\Delta t} \,  \Gamma_{ k}(\tau)\, d\tau = \Delta t \Gamma_{ k}(t)  + \frac{\Delta t^2}{2} \Gamma_{ k}'(\xi) \\
=  \Delta t \Gamma_{ k}(t) \left(1 + \frac{\Delta t \Gamma_{ k}'(\xi)}{2 \Gamma_{ k}(t) }  \right),
\end{align*}
where  $\xi \in [t, t+\Delta t]$. Therefore, we should always keep 
$\frac{\Delta t |\Gamma'_{ k,max}|}{2 \Gamma_{ k}(t) } $ small, with $\Gamma'_{ k,max}$  the maximum of $|\Gamma_{ k}' |$  in $[t, t+\Delta t]$.

 The obvious difficulty is to control the maximum of $|\Gamma_{ k}' |$  in the interval $[t, t+\Delta t]$, in which we have not evolved yet (we are still in time $t$). The best we can do is simply to estimate its values, for instance, $ \Gamma_k'(t)$ can be estimated  at time $t=t_i$ by the finite difference formula 
$ \Gamma_{k,\rm approx}' (t_i) = \frac{\Gamma_k(t_i) - \Gamma_k(t_{i-1})}{t_i - t_{i-1}}  $ 
and check after the evolution at time $t_{i+1} = t+\Delta t$ if we were correct or not.
But how to be sure that  $\Gamma_{ k}$ will not change abruptly within $[t, t+\Delta t]$, creating a much bigger error than anticipated?
There is no universal solution to this problem. The only way is to rely on our knowledge of the typical evolution of the rates to avoid such situations. In our case, the rates evolve smoothly and are typically of Gaussian, Lorentzian, or exponential shape.

A good illustration of the problem is given by the exponential case. For instance, for an exponential growth (like our ionizing rate)  
$\Gamma_k(t) = A e^{t/\tau}$. So, in $[t, t+\Delta t]$ the maximum of $|\Gamma_{ k}' |$ is given  by $\Gamma'_k(t)   e^{ \Delta t/\tau}$. Clearly an estimation of $\Gamma'_{ k,max}$ by  $\Gamma'_k(t)$ is correct only if $ \frac{\Delta t}{\tau}   \ll 1$. But because, in this exponential example, $ \frac{\Delta t}{\tau} = \frac{\Delta t \Gamma'_k(t) }{\Gamma_k(t)}$ we see that this condition is already $ \Delta t \Gamma'_k(t) \ll {\Gamma_k(t)}$. This also indicates that using the second order (such as used in trapezoidal   or Simpson's rules for instance)
$
\int_t^{t+\Delta t} \, \Gamma_{ k}(\tau)\, d\tau = \Delta t \Gamma_{ k}(t)  + \frac{\Delta t^2}{2} \Gamma_{ k}'(t) + \frac{\Delta t^3}{6} \Gamma_{ k}''(\xi)
$
where  $\xi \in [t, t+\Delta t]$,
will not greatly improve the result because our condition indicator $\frac{\Delta t^2}{2} \Gamma_{ k}'(t)$ will be negligible in front of $ \Delta t \Gamma_{ k}(t)  $. Thus, using the exponential case illustrates that, regardless of the methods used, the error always seems to be controlled by $\frac{\Delta t \Gamma'_k(t)}{\Gamma_k(t)}$.

We can indeed check that for Gaussian, Lorentzian or exponential $\Gamma_k(t)$ functions  using  $\Delta t$  smaller that  $ \Delta t_{\rm max}(t) = \epsilon \frac{ \Gamma_k(t) }{|\Gamma'_k(t)|}  $, with $0< \epsilon <1$, ensure the error to be smaller than $\epsilon$, in a sens that we always have
\begin{equation}
 \frac{\left| \int_t^{t+\Delta t} \,  \Gamma_{ k}(\tau)\, d\tau  - \Delta t \Gamma_{ k}(t) \right| }{\Delta t \Gamma_{ k}(t) } \leq \epsilon \label{error}
\end{equation}

After this discussion, we can thus safely choose the maximum $\Delta t$ compatible with the bounding error we want by choosing 
$ \Delta t_{\rm max}(t) = \epsilon \frac{ \Gamma_k(t) }{|\Gamma'_k(t)| }  $. 

However we cannot calculate precisely $|\Gamma'_k(t)|$ but only evaluate it for instance by choosing
$ \Delta t_{\rm max,approx}(t_i) = \epsilon \frac{ \Gamma_k(t_i) }{|\Gamma_{k,\rm approx}' (t_i) | }  $.  It is less obvious to control the error using this approximate formula because it depends on the previous choices for the time steps $t_i-t_{i-1}$. However,  this timestep evolves iteratively not too far from  $ \Delta t_{\rm max}(t_{i-1})$. Indeed, we have verified that, by running the algorithm (defined below), the error (\ref{error}) remains correct for $\epsilon < 0.3$ in the cases of Gaussian, Lorentzian, or exponential rate equations. 

 Let's mention a practical difficulty that exists for all algorithms if the system also evolves dynamically under another independent timing. For instance, in our case, or in any molecular dynamics simulation, the trajectory evolution under external forces, which imposes such dynamical time ($dt_{\rm SIMION}$ in our case). One simple solution to combine the external degree of freedom evolution (the trajectories) and the internal one (the states evolution) is to evolve the master equation using the shortest time between the one used for trajectory evolution and the one required to make the state evolution accurate enough. 

Thus, we are now ready to propose our algorithm. It is essentially the same as the one proposed in \cite{prados1997dynamical} except that the evolving time $\Delta t$ is not fixed (it is actualized every step number $i$). Furthermore, it is bounded by the dynamical time ($dt_{\rm SIMION}$ for us) and, maybe more importantly, the error is controlled by $ \epsilon $, with $0< \epsilon <0.3$. 

In brief, the essence of the algorithm is the following: we note $I$ (to be understood as $I_i$) our current estimation of $\int_{t_0}^{t_i}  \Gamma_{ k} (\tau) d \tau$. Thus,  the  reaction  time $t' =t_i + \Delta t$ given by
$ - \ln u = \int_{t_0}^{t'}  \Gamma_{ k} (\tau) d \tau $ can be estimated by
	$ - \ln u = I +   \Gamma_{ k} (t_i) \Delta t  $.
     
\begin{enumerate}

    \item At time $t=t_0$, the system has evolved (or has been created) in a given state  $k$. 
We choose a unit-interval uniform random number generator  $u$: $0<u\leq 1$. 
We initiate $i=0$, $I=0$. The safer is to initiate  $ \Gamma_{k,\rm approx}'(t_0)$ to a very large value so that the first time step ($\Delta t_{\rm max,approx}(t_0)$) will be small enough to ensure proper estimation of $ \Gamma_{k,\rm approx}'$, but if the evolution is smooth enough it is also possible to initiate $ \Gamma_{k,\rm approx}'(t_0) =0$.

	\item Calculate the new rate list $ \Gamma_{l\leftarrow k} (t_i) $ for all possible states $|l\rangle$, calculate $\Gamma_{ k}(t_i)  = \sum_{l=1}^N \Gamma_{l\leftarrow k} (t_i)$ and estimate $ \Gamma_k'(t_i) $  by (for $i>0$) $ \Gamma_{k,\rm approx}'(t_i) =  \frac{\Gamma_k(t_i) - \Gamma_k(t_{i-1})}{t_i - t_{i-1}}  $.
    
	\item  Take for the evolution time $\Delta t $, the smallest of the three times (the first 2 being skipped, or taken as large values, if $\Gamma_k(t_i) =0$): 
    
    \begin{itemize} 
    \item $\Delta t_1$ the one given by $ \Delta t = \frac{- \ln u - I}{   \Gamma_{ k} (t_i) }$.

\item $\Delta t_2$ the maximum allowed one, to control the error $\epsilon$, 
     $ \Delta t_{\rm max,approx}(t_i) = \epsilon \frac{ \Gamma_k(t_i) }{|\Gamma_{k,\rm approx}'(t_i)| }  $ (care has to be done to avoid division by zero, either with a  test,  either by adding a small offset in the denominator)
     
    \item $\Delta t_3$ the dynamical  time ($dt_{\rm SIMION}$ for us).
     
     \end{itemize}
     Put $t_{i+1} = t_i + \Delta t$ and evolve the system by forcing $dt_{\rm SIMION}$ to be the time found. 
     
    \item If the time chosen  $\Delta t $ was the first one (so $\Delta t_1$),  goes to the next step (step 5), if not make increase $I$ by $\Delta t \Gamma_{ k} (t_{i})  $,  increase $i$ by one unit, and go back to step 2.
    
	\item We are at time $t'=t_{i+1}$ and so a reaction occur. We
     calculate the new rate list $ \Gamma_{l k} (t') $. We choose a unit-interval uniform random number generator  $u'$: $0<u'\leq 1$ and searching for the integer $l$ for which $R_{l - 1} < u' R_N \leq  R_l$ where
 $R_j = \sum_{i=1,j} \Gamma_{i k} (t')$ and $R_0=0$. 
 
	\item Setting the system to state $|l\rangle$. Then go back to step 1.
\end{enumerate}


\bibliographystyle{h-physrev}
\bibliography{biblio}

\begin{thebibliography}{10}

\bibitem{Materials2019}
{\em Materials Characterization} (ASM International, 2019).

\bibitem{laucht2021roadmap}
A.~Laucht {\em et~al.},
\newblock Nanotechnology {\bf 32}, 162003 (2021).

\bibitem{hoflich2023roadmap}
K.~H\"{o}flich {\em et~al.},
\newblock Applied Physics Reviews {\bf 10}, 041311 (2023).

\bibitem{hawkes2019springer}
P.~W. Hawkes and J.~C. Spence,
\newblock {\em Springer handbook of microscopy} (Springer Nature, 2019).

\bibitem{li2021recent}
P.~Li {\em et~al.},
\newblock Nanoscale {\bf 13}, 1529 (2021).

\bibitem{Gallagher.1974}
A.~C. Gallagher and G.~York,
\newblock Review of Scientific Instruments {\bf 45}, 662 (1974).

\bibitem{musumeci2018advances}
P.~Musumeci {\em et~al.},
\newblock Nuclear Instruments and Methods in Physics Research Section A:
  Accelerators, Spectrometers, Detectors and Associated Equipment {\bf 907},
  209 (2018).

\bibitem{mcclelland2016bright}
J.~J. McClelland {\em et~al.},
\newblock Applied physics reviews {\bf 3} (2016).

\bibitem{mcculloch2016cold}
A.~J. McCulloch, B.~M. Sparkes, and R.~E. Scholten,
\newblock Journal of Physics B: Atomic, Molecular and Optical Physics {\bf 49},
  164004 (2016).

\bibitem{gallagher1994rydberg}
T.~F. Gallagher,
\newblock Rydberg atoms,
\newblock in {\em Springer Handbook of Atomic, Molecular, and Optical Physics},
  pp. 231--240, Springer, 1994.

\bibitem{hahn2021cesium}
R.~Hahn, A.~Trimeche, C.~Lopez, D.~Comparat, and Y.~Picard,
\newblock Physical Review A {\bf 103}, 042821 (2021).

\bibitem{trimeche2020ion}
A.~Trimeche, C.~Lopez, D.~Comparat, and Y.~Picard,
\newblock Physical Review Research {\bf 2}, 043295 (2020).

\bibitem{lopez2019real}
C.~Lopez, A.~Trimeche, D.~Comparat, and Y.~Picard,
\newblock Physical Review Applied {\bf 11}, 064049 (2019).

\bibitem{mcculloch2017field}
A.~McCulloch {\em et~al.},
\newblock Physical Review A {\bf 95}, 063845 (2017).

\bibitem{viteau2016ion}
M.~Viteau {\em et~al.},
\newblock Ultramicroscopy {\bf 164}, 70 (2016).

\bibitem{antoni2018watt}
L.~Antoni-Micollier {\em et~al.},
\newblock Optics letters {\bf 43}, 3937 (2018).

\bibitem{reveillard2018coldfib}
M.~Reveillard {\em et~al.},
\newblock Microscopy and Microanalysis {\bf 24}, 804 (2018).

\bibitem{hahn2022comparative}
R.~Hahn {\em et~al.},
\newblock Review of Scientific Instruments {\bf 93} (2022).

\bibitem{manura20088}
D.~Manura and D.~A. Dahl,
\newblock Simion 8.0 user manual; scientific instrument services, inc: Ringoes,
  nj, usa, 2008.

\bibitem{viray2021photoionization}
M.~A. Viray, E.~Paradis, and G.~Raithel,
\newblock New Journal of Physics {\bf 23}, 063022 (2021).

\bibitem{comparat2014molecular}
D.~Comparat,
\newblock Physical Review A {\bf 89}, 043410 (2014).

\bibitem{moufarej2017forced}
E.~Moufarej {\em et~al.},
\newblock Physical Review A {\bf 95}, 043409 (2017).

\bibitem{chardonnet1984generation}
C.~Chardonnet, D.~Delande, and J.~Gay,
\newblock Optics communications {\bf 51}, 249 (1984).

\bibitem{Harmin1984PRA}
D.~A. Harmin,
\newblock Phys. Rev. A {\bf 30}, 2413 (1984).

\bibitem{ceccatto1986effective}
H.~Ceccatto,
\newblock Physical Review B {\bf 33}, 4734 (1986).

\bibitem{prados1997dynamical}
A.~Prados, J.~Brey, and B.~S{\'a}nchez-Rey,
\newblock Journal of statistical physics {\bf 89}, 709 (1997).

\bibitem{kime2013high}
L.~Kime {\em et~al.},
\newblock Physical Review A: Atomic, Molecular, and Optical Physics {\bf 88},
  033424 (2013).

\bibitem{reveillard2022efficient}
M.~Reveillard {\em et~al.},
\newblock The European Physical Journal D {\bf 76}, 35 (2022).

\bibitem{xie2022cold}
W.~Xie {\em et~al.},
\newblock AIP Advances {\bf 12} (2022).

\bibitem{mitchell2025selective}
K.~T. Mitchell {\em et~al.},
\newblock Phys. Rev. A {\bf 112}, 023108 (2025).

\bibitem{grimm2000optical}
R.~Grimm, M.~Weidem{\"u}ller, and Y.~B. Ovchinnikov,
\newblock Optical dipole traps for neutral atoms,
\newblock , Advances In Atomic, Molecular, and Optical Physics Vol.~42, pp.
  95--170, Academic Press, 2000.

\bibitem{fichthorn1991theoretical}
K.~A. Fichthorn,
\newblock The Journal of chemical physics {\bf 95}, 1090 (1991).

\bibitem{binder2019monte}
K.~Binder and D.~W. Heermann,
\newblock {\em Monte Carlo Simulation in Statistical Physics: An
  Introduction}Graduate Texts in Physics, 6 ed. (Springer Cham, 2019),
\newblock 150 b/w illustrations, 5 illustrations in colour.

\bibitem{landau2021guide}
D.~Landau and K.~Binder,
\newblock {\em A guide to Monte Carlo simulations in statistical physics}
  (Cambridge university press, 2021).

\bibitem{jansen2003introduction}
A.~P.~J. Jansen,
\newblock {\em An introduction to kinetic Monte Carlo simulations of surface
  reactions} (Springer, Berlin, Germany, 2012).

\bibitem{bortz1975new}
A.~B. Bortz, M.~H. Kalos, and J.~L. Lebowitz,
\newblock Journal of Computational physics {\bf 17}, 10 (1975).

\bibitem{gillespie1977exact}
D.~T. Gillespie,
\newblock The journal of physical chemistry {\bf 81}, 2340 (1977).

\bibitem{anderson2007modified}
D.~F. Anderson,
\newblock The Journal of chemical physics {\bf 127} (2007).

\bibitem{thanh2015simulation}
V.~H. Thanh and C.~Priami,
\newblock The Journal of chemical physics {\bf 143} (2015).

\bibitem{andersen2019practical}
M.~Andersen, C.~Panosetti, and K.~Reuter,
\newblock Frontiers in chemistry {\bf 7}, 202 (2019).

\bibitem{chittari2024revisiting}
S.~S. Chittari and Z.~Lu,
\newblock The Journal of Chemical Physics {\bf 161} (2024).

\end{thebibliography}

\end{document}